\documentclass[cmp]{svjour}  
\usepackage{amsmath}
\usepackage{amsfonts,amssymb}
\usepackage[vcentermath]{youngtab}

\begin{document}

\title{The Pauli principle revisited}

\author{ Murat Altunbulak and Alexander Klyachko}
\institute{Department of Mathematics, Bilkent University, 06800 Bilkent, Ankara, Turkey.\\
\email{murata@fen.bilkent.edu.tr; klyachko@fen.bilkent.edu.tr}}

\maketitle
\begin{abstract} By the Pauli exclusion principle  no
quantum state can be occupied by more than one electron. One can
put it as a constraint on the electron density matrix that bounds
its eigenvalues by $1$. Shortly after its discovery the Pauli
principle has been replaced by skew symmetry of a multi-electron
wave function. In this paper we solve a longstanding problem about
the impact of this replacement on the electron density matrix,
that goes far beyond the original Pauli principle.
\end{abstract}
\tableofcontents

\section{Introduction}
The Pauli exclusion principle, discovered in 1925, claims that no
quantum state can be occupied by more than one electron. In terms
of the electron density matrix\footnote{There is no agreement on a
proper normalization of the one-electron matrix. To avoid a
confusion we call it {\it electron density\/} for Dirac's
normalization to the number of particles
$\mathop{\mathrm{Tr}}\rho=N$, and reserve the term {\it reduced
state\/} for the probability normalization
$\mathop{\mathrm{Tr}}\rho=1$.}
$\rho$ this amounts to the inequality
$\langle\psi|\rho|\psi\rangle\le 1$, that bounds  its eigenvalues
by one.  The following year Heisenberg and Dirac replaced the
Pauli principle by skew symmetry of a multi-electron wave function
\cite[Ch. 4]{Duck}.

The subject of this study is the impact of this replacement on the
electron density matrix. The latter determines the light
scattering and  therefore quite literally represents  a visible
state of the electron system. The impact goes far beyond the
original Pauli principle. As an example, consider three electron
system $\wedge^3\mathcal{H}_6$ with one-electron space
$\mathcal{H}_6$ of dimension $6$.
Then the spectrum $\lambda$ of the electron density matrix,
arranged in non-increasing order, is bounded by the following
(in)equalities discovered by Borland and Dennis \cite{Borland
Dennis}
\begin{equation}\label{B-D1}
\lambda_1+\lambda_6=\lambda_2+\lambda_5=\lambda_3+\lambda_4=1,\quad
\lambda_4\le \lambda_5+\lambda_6.
\end{equation}
The authors established  the sufficiency of these constraints and
referred for
a complete proof to M.B.~Ruskai and
R.L.~Kingsley.\footnote{Recently M.B.~Ruskai published the proof
\cite{Ruskai07} derived
from known constraints on the spectra of Hermitian matrices $A$,
$B$, and $C=A+B$.
Conceptually the $N$-representability problem is close to  the
Hermitian spectral problem \cite{Kl98,Kl02}, but a direct
connection between them, beyond sporadic coincidences,  is
unlikely. An independent R.L.~Kingsley's solution apparently  has
never been published.}
It worth reading their comment:
\begin{quote}
{\it We have no apology for consideration of such a special case.
The general $N$-representability problem is so difficult and yet
so fundamental for many branches  of science that each concrete
result is useful in shedding light on the nature of general
solution.}
\end{quote}
In spite of some bogus claims \cite{Peltzer}, refuted in
\cite{Ruskai}, this result had stood for more then three decades
as the only known solution of the $N$-representability problem
beyond two electrons $\wedge^2\mathcal{H}_r$ and two holes
$\wedge^{r-2}\mathcal{H}_r$.
For the latter systems the problem is easy
and the constraints amounts  to double degeneracy of the spectrum,
starting from the head
$\lambda_{2i-1}=\lambda_{2i}$ for two electrons
and from the tail $\lambda_{r-2i}=\lambda_{r-2i-1}$ for two holes
\cite{Coleman63}, where we set $\lambda_i=0$ for $i>r$, and
$\lambda_i=1$ for $i<1$.

Here we solve
this longstanding problem.
The content of the paper is as follows.

In Section \ref{digest} we recast the Berenstein-Sjamaar theorem
\cite[Thm 3.2.1]{Ber-Sja} into a usable form
(Theorem \ref{Ber_Sja}). This provides a theoretical basis for our
study.

We start Section \ref{H_lambda} by a variation of the above
problem, called {\it $\nu$-representabi\-lity\/}, that takes into
account both spin and orbital occupation numbers. Mathematically
this amounts to replacing the exterior power $\wedge^N\mathcal{H}$
by a representation $\mathcal{H}^\nu$ defined by Young diagram
$\nu$ of order $N$. Theorem \ref{nu thm} gives a formal solution
of the $\nu$-representability problem. We derive from it the
majorization inequality $\lambda\preceq\nu$, that plays the
r\^{o}le of the Pauli principle. This inequality is necessary and
sufficient for $\lambda$ to be occupation numbers of an
unspecified {\it mixed\/} state (Theorem \ref{mixed_occ}). Theorem
\ref{mlt_clmn} deals with a
class of systems where the majorization inequality alone provides a
criterion for {\it pure\/} $\nu$-representability. This includes the
so-called {\it closed shell\/}, meaning
a system of electrons of total spin zero. The corresponding Young
diagram $\nu$ consists of two columns of equal length. For this
system all constraints on the occupation numbers are given by the
Pauli type inequality $\lambda\le 2$. In the next Theorem
\ref{nu_coeff} we calculate the topological coefficients
$c_w^v(a)$ that governed the constraints on the occupation numbers
in Theorem \ref{nu thm}. This gives  it the full strength we need
in the next section.

Section \ref{beyond} starts with analysis of pure
$\nu$-representability for a toy example of two-row diagrams, that
allows us to illustrate the basic technique (Theorem
\ref{two_row_thm}). These are exceptional systems where the
constraints on the occupation numbers are given by a {\it
finite\/} set of inequalities independent of the rank.
Then we return to the original $N$-representability problem, that
appears to be the most difficult one. For example, in contrast to
Theorem \ref{two_row_thm}, no finite system of inequalities can
describe $N$-representability for a fixed $N>1$ and arbitrary big
rank (Corollary \ref{no_fnt} to Proposition \ref{row_diag}). This
forces us to restrict either the rank, as we do in the last
section, or the type of the inequalities. Here we focus on the
inequalities with $0/1$ coefficients. It turns out that under some
natural conditions such an inequality should be either of the form
\begin{equation}\label{1st}
\lambda_{i_1}+\lambda_{i_2}+\cdots+\lambda_{i_{N-1}}\le N-2,
\end{equation}
with $\sum_k (i_k-k)=r-N+1$, or of the form
\begin{equation}\label{2nd}
\lambda_{i_1}+\lambda_{i_2}+\cdots+\lambda_{i_p}\le N-1,
\end{equation}
with $p\ge N$ and $\sum_k (i_k-k)=\binom{p}{N}$. We call them {\it
Grassmann inequalities\/} of the first and second kind
respectively. A surprising result is that these inequalities
actually hold true with very few exceptions (Theorems
\ref{Gr_kind2} and \ref{kind2_thm}).

In the simplest case  $N=3$ we get from (\ref{1st}) inequalities
$$\lambda_{k+1}+\lambda_{r-k}\le 1,\quad 0\le k<(r-1)/2$$
that  hold for any even rank $r\ge6$.  This constraint prohibits
more than one electron to occupy {\it two\/} symmetric orbitals
and supersedes the original Pauli principle. For $r=6$, due to the
normalization $\sum_i\lambda_i=3$, the inequalities degenerate
into Borland-Dennis {\it equalities\/} (\ref{B-D1}). For odd rank
the first inequality $k=0$ should be either skipped or replaced by
weaker one $\lambda_1+\lambda_r\le 1+\frac{2}{r-1}$.

We treat Grassmann inequalities of the second kind (\ref{2nd}) only
for lowest levels $p=N,N+1$. For $N=3$ and $p=N+1$ they amount to
four inequalities
\begin{equation}\label{rank7}
\begin{array}{ll}
\quad \lambda_2+\lambda_3+\lambda_4+\lambda_5\le 2,&\qquad\quad
\lambda_1+\lambda_3+\lambda_4+\lambda_6\le2,\\
\quad \lambda_1+\lambda_2+\lambda_5+\lambda_6\le 2,& \qquad\quad
\lambda_1+\lambda_2+\lambda_4+\lambda_7\le2,
\end{array}
\end{equation}
that hold for arbitrary rank $r$ and give all the constraints
for $r\le 7$. For $r=6$ they turn into Borland-Dennis conditions
(\ref{B-D1}).

In the next Section \ref{repr_thry} we briefly  discuss a
connection of the $\nu$-representability with representation
theory, that provides information complementary to Theorem~\ref{nu
thm}. A combination of the two approaches  leads to an algorithm
for solution of the problem for any fixed rank. The algorithm,
along with other tools, has been used in calculations reported in
the last Section \ref{small}. Eventually
this led to a complete solution
of the $N$-representability problem for rank $r\le 10$. However,
we provide a rigorous justification only for $r\le 8$. We also
give an example of constraints on the spin and orbital occupation
numbers for a system of three electrons of total spin $1/2$.

The first sections may be mathematically more demanding then the
rest of the paper. We recommend books \cite{Ful_Harr,Ful97,Ful98}
as a general reference on Schubert calculus, Lie algebra, and
representation theory.

The theoretical results of the paper belong to the second author.
They were often inspired by calculations, that at this stage
couldn't be accomplished by a computer without intelligent human
assistance and insight.
\section{A digest of the Berenstein-Sjamaar paper}\label{digest}
Let $M$ be a compact connected Lie group with the Lie algebra
$\mathfrak{m}$ and its dual coadjoint representation
$\mathfrak{m}^*$.
For coadjoint orbit $\mathcal{O}\subset\mathfrak{m}^*$ of group
$M$ and a Cartan subalgebra $\mathfrak{t}\subset\mathfrak{m}$
consider the composition
$\Delta:\mathcal{O}\hookrightarrow\mathfrak{m}^*\rightarrow\mathfrak{t}^*$
known as the {\it moment map\/}. By Kostant's theorem its image is
a convex polytope spanned by the $W$-orbit of some weight $\mu\in
\mathfrak{t}^*$ which can be taken from a fixed positive Weyl
chamber $\mathfrak{t}^*_+$. Here
$W=N(\mathfrak{t})/Z(\mathfrak{t})$ is the Weyl group of $M$. This
gives a parameterization of the coadjoint orbits $\mathcal{O}_\mu$
by the dominant weights $\mu\in
\mathfrak{t}^*_+$.
\begin{example}\label{U(n)} In this paper we will mostly deal with the unitary
group $\mathrm{U}(n)$ whose Lie algebra $\mathfrak{u}(n)$ consists
of all Hermitian\footnote{\label{Herm}Hereafter  we treat
$\mathfrak{u}(n)$ as the algebra of Hermitian, rather than
skew-Hermitian, operators at the expense of a modified Lie bracket
$[X,Y]=i(XY-YX)$.} $n\times n$ matrices. Let us identify
$\mathfrak{u}(n)$ with its dual via the invariant trace form
$(A,B)=\mathrm{Tr}(AB)$. Then the (co)adjoint orbit
$\mathcal{O}_\mu$ consists of all Hermitian matrices $A$ of
spectrum $\mu:\mu_1\ge\mu_2\ge\cdots\ge\mu_n$ and the moment map
$\Delta:\mathcal{O}_\mu\rightarrow \mathfrak{t}$ is given by
orthogonal projection into the Cartan subalgebra of diagonal
matrices $\mathfrak{t}$. Kostant's theorem in this case amounts to
Horn's observation that the diagonal entries of Hermitian matrices
of spectrum $\mu$ form a convex polytope with vertices $w\mu$
obtained from $\mu$ by permutations
of the coordinates $\mu_i$. This is equivalent to the {\it
majorization inequalities\/}
\begin{eqnarray}d_1&\le&\mu_1\nonumber\\
d_1+d_2&\le&\mu_1+\mu_2\nonumber\\
d_1+d_2+d_3&\le&\mu_1+\mu_2+\mu_3\label{maj_inq}\\
\cdots&\cdots&\cdots\nonumber\\
d_1+d_2+\cdots+d_n&=&\mu_1+\mu_2+\cdots+\mu_n\nonumber
\end{eqnarray} for the diagonal entries $d:d_1\ge d_2\ge \cdots\ge
d_n$ of matrix $A$. We will use for them a shortcut $d\preceq\mu$.

\end{example}

Consider now  an immersion $f:L\rightarrow M$ of another compact
Lie group $L$ and the induced morphisms
$f_*:\mathfrak{l}\hookrightarrow\mathfrak{m}$ and
$f^*:\mathfrak{m}^*\rightarrow\mathfrak{l}^*$ of the Lie algebras
and their duals.
In the paper \cite{Ber-Sja} Berenstein and Sjamaar
found  a decomposition of the projection
$f^*(\mathcal{O}_\mu)\subset \mathfrak{l}^*$ of an $M$-orbit
$\mathcal{O}_\mu\subset \mathfrak{m}^*$ into $L$-orbits
$\mathcal{O}_\lambda\subset f^*(\mathcal{O}_\mu)$. Here we
paraphrase their main result in the form suitable for the intended
applications.

Fix a Cartan subalgebras
$\mathfrak{t}_L\hookrightarrow\mathfrak{t}_M$
and for every {\it test spectrum\/} $a\in \mathfrak{t}_L$ consider
the inclusion of the adjoint orbits of groups $L$ and $M$
\begin{equation}\label{phi_a}\varphi_a:\mathcal{O}_a\hookrightarrow\mathcal{O}_{f_*(a)}
\end{equation}
through $a$ and $f_*(a)$ respectively.
Topologically the orbits are (generalized) {\it flag varieties\/}.
They carry a hidden complex structure coming from the
representation
\begin{equation}\label{P_a}
\mathcal{O}_a=L/Z_L(a)=L^\mathbb{C}/P_a
\end{equation} where $P_a\subset
L^\mathbb{C}$ is a parabolic subgroup of the complexified group
$L^\mathbb{C}$ whose Lie algebra $\mathfrak{p}_a$ is spanned by
$\mathfrak{t}_L$ and the root vectors $X_\alpha$ such that
$\langle\alpha,a\rangle\ge 0$. One can say this in another way
$$P_a=\{g\in
L^\mathbb{C}\mid\lim_{t\rightarrow-\infty}e^{ta}ge^{-ta}\mbox{
exists}\}$$ which makes it clear that $f:P_a\rightarrow
P_{f_*(a)}$.

We will use the parabolic subgroups  to construct {\it canonical
bases\/} in cohomologies $H^*(\mathcal{O}_a)$ and
$H^*(\mathcal{O}_{f_*(a)})$.
Let $T_L\subset B\subset P_a$ be a Borel subgroup containing a
maximal torus $T_L$ with Lie algebra $\mathfrak{t}_L$.
The flag variety $\mathcal{O}_a=L^\mathbb{C}/P_a$ splits into
disjoint union of {\it Schubert cells\/}  $Bv P_a/P_a$,
parameterized by
the left cosets $v\in W_L/W_{Z_L(a)}$ or in practice by
representatives of minimal length $\ell=\ell(v)$ in these cosets.
We actually prefer to deal with shifted cells
$v^{-1}BvP_a/P_a=B^vP_a/P_a$ depending on the Borel subgroups
$B^v\supset T_L$ modulo conjugation by the Weyl group of the
centralizer $W(Z_L(a))$. The closure of $B^vP_a/P_a$ is known as
the {\it Schubert variety\/}, and its cohomology class
$\sigma_v\in H^{2\ell(v)}(\mathcal{O}_a)$ is called the {\it
Schubert cocycle\/}. These cocycles form the {\it canonical
basis\/} of the cohomology ring $H^*(\mathcal{O}_a)$.



Inclusion (\ref{phi_a}) induces a morphism of the cohomologies
\begin{equation}\label{cohom_phi}\varphi_a^*:H^*(\mathcal{O}_{f_*(a)})\rightarrow
H^*(\mathcal{O}_a),
\end{equation}
given in the canonical bases by the coefficients $c_w^v(a)$ of the
decomposition
\begin{equation}\label{coeff}
\varphi_a^*:\sigma_w\mapsto \sum_v c_w^v(a)\sigma_v.
\end{equation}
They play a crucial r\^{o}le in the next theorem. We
extend them by zeros if either $v\in W_L$ or $w\in W_M$ is not the
minimal representative of a coset in $W_L/W_{Z_L(a)}$ or
$W_M/W_{Z_M(f_*(a))}$ respectively.


\begin{theorem}\label{Ber_Sja} In the above notations the inclusion $\mathcal{O}_\lambda\subset
f^*(\mathcal{O}_\mu)$ is equivalent to the following system of
linear inequalities
\begin{equation}\label{mod_BS}
\langle\lambda,va\rangle\le\langle\mu,wf_*(a)\rangle\tag{$a,v,w$}
\end{equation} for all $a\in\mathfrak{t}_L,v\in W_L,w\in W_M$ such that $c_w^v(a)\ne 0$.
\end{theorem}
\proof
This is not the way how Berenstein and Sjamaar stated their
result. Instead, for some generic $a_0\in \mathfrak{t}_L$ they fix
positive Weyl chambers $\mathfrak{t}_L^+\ni a_0$ and
$\mathfrak{t}_M^+\ni f_*(a_0)$ and use them to define Schubert
cocycles $\sigma_v\in H^*(\mathcal{O}_a)$ and $\sigma_w\in
H^*(\mathcal{O}_{f_*(a)})$ for all other $a\in \mathfrak{t}_L^+$.
Hence their Schubert cocycles $\sigma_w$ are canonical in the
above sense iff $f_*(a)$ and $f_*(a_0)$ are in the same Weyl
chamber. The set of such $a\in \mathfrak{t}_L^+$ form a convex
polyhedral cone  called the {\it principle cubicle\/}. It is
determined by $a_0$, and different choices of $a_0$ produce a
polyhedral decomposition of the positive Weyl chamber
$\mathfrak{t}_L^+$ into cubicles.

For every cubicle Berenstein and Sjamaar  gave a system of linear
constraints on the dominant weights $\lambda,\mu$, so that all
together they
provide a criterion for the inclusion $\mathcal{O}_\lambda\subset
f^*(\mathcal{O}_\mu)$. For the principal cubicle
the constraints are most simple and  look as follows \cite[Thm
3.2.1]{Ber-Sja}
\begin{equation}\label{B-S.3.5}
v^{-1}\lambda\in f^*(w^{-1}\mu-\mathcal{C}),\quad \mbox{ for }
\quad c_w^v(a_0)\ne 0,
\end{equation} where $\mathcal{C}$ is
a cone spanned by the positive roots in $\mathfrak{t}_M^*$. Note
that $f^*(\mathcal{C})$ is the cone dual to the principal cubicle
and therefore  the above condition
can be recast  into  the inequalities
\begin{equation}
\langle v^{-1}\lambda,a\rangle\le \langle
f^*(w^{-1}\mu),a\rangle\Longleftrightarrow \langle
\lambda,va\rangle\le \langle \mu,wf_*(a)\rangle,
\end{equation}
that hold for all $a$ from the principle cubicle {\it provided\/}
that $c_w^v(a_0)\ne 0$. The coefficients $c_w^v(a)$ are actually
constant inside the cubicle, and therefore the last condition can
be changed to $c_w^v(a)\ne 0$. Thus we arrived at the inequalities
(\ref{mod_BS}) for the principle cubicle. Other inequalities
(\ref{mod_BS}) follow by choosing another cubicle as the principle
one. They are equivalent to the remaining more complicated
inequalities in \cite[Thm 3.2.1]{Ber-Sja}, but look differently
since Berenstein and Sjamaar use other non-canonical Schubert
cocycles. \qed
\begin{example} {\it Quantum marginal problem\/ {\rm \cite{Klyachko2004}}}.\label{quant_mrg}
Let's illustrate the above theorem with immersion of unitary
groups
$$f:\mathrm{U}(\mathcal{H}_A)\times
\mathrm{U}(\mathcal{H}_B)\rightarrow \mathrm{U}(\mathcal{H}_{AB}),\qquad
g_A\times g_B\mapsto g_A\otimes g_B,$$ where
$\mathcal{H}_{AB}=\mathcal{H}_A\otimes\mathcal{H}_B$. As we have
seen in Example \ref{U(n)} the coadjoint orbit of
$\mathrm{U}(\mathcal{H}_{AB})$
consists of the isospectral Hermitian operators
$\rho_{AB}:\mathcal{H}_{AB}$ understood here as {\it mixed
states\/}. The projection
$$f^*(\rho_{AB})=\rho_A\otimes 1+1\otimes \rho_B$$
amounts to {\it reduced operators\/} $\rho_A:\mathcal{H}_A$ and
$\rho_{B}:\mathcal{H}_B$ implicitly defined by the equations
\begin{equation}\label{reduced}
\mathrm{Tr}_{\mathcal{H}_{A}}(\rho_{A}X_A)=\mathrm{Tr}_{\mathcal{H}_{AB}}(\rho_{AB}X_A),
\qquad
\mathrm{Tr}_{\mathcal{H}_{B}}(\rho_{A}X_B)=\mathrm{Tr}_{\mathcal{H}_{AB}}(\rho_{AB}X_B)
\end{equation} for all Hermitian operators $X_A:\mathcal{H}_A$ and $X_B:\mathcal{H}_B$.
This means that $\rho_A$, $\rho_B$ are just the visible states of
the subsystems $\mathcal{H}_A$, $\mathcal{H}_B$. In this settings
Theorem \ref{Ber_Sja} tells that all constraints on the decreasing
spectra $\lambda^{AB}=\text{Spec}(\rho^{AB})$,
$\lambda^{A}=\text{Spec}(\rho^{A})$, and
$\lambda^{B}=\text{Spec}(\rho^{B})$ are given by the inequalities
\begin{equation}\label{MargIneq}
\sum_i a_i\lambda^A_{u(i)}+\sum_j b_j\lambda^B_{v(j)}\le \sum_{k}
(a+b)^\downarrow_k\lambda^{AB}_{w(k)},
\end{equation}
for all test spectra $a:a_1\ge a_2\ge \cdots\ge a_n$, $b:b_1\ge
b_2\ge\cdots\ge b_m$ from the Cartan subalgebras $\mathfrak{t}_A$,
$\mathfrak{t}_B$ and permutations $u,v,w$ such that
$c_{w}^{uv}(a,b)\ne 0$. Here $(a+b)^\downarrow$ denotes the
sequence $a_i+b_j$ arranged in decreasing order. The order
determines the canonical Weyl chamber containing $f_*(a,b)$. The
pairs $(a,b)$ with fixed order of terms $a_i+b_j$ in
$(a+b)^\downarrow$ form a cubicle.

The adjoint orbit $\mathcal{O}_a\subset
\mathfrak{u}(\mathcal{H}_A)$ is a classical flag variety
understood as the set of Hermitian operators $X_A:\mathcal{H}_A$
of spectrum $a=\mathop{\mathrm{Spec}}X_A$. Denote it by
$\mathcal{F}_a(\mathcal{H}_A)$. Then the morphism (\ref{phi_a}) is
given by the equation
\begin{equation}
\varphi_{ab}:\mathcal{F}_a(\mathcal{H}_A)\times
\mathcal{F}_b(\mathcal{H}_B)\rightarrow
\mathcal{F}_{a+b}(\mathcal{H}_{AB}),\quad (X_A,X_B)\mapsto
X_A\otimes 1+1\otimes X_B
\end{equation}
and the coefficients $c^{uv}_w(a,b)$ are determined by the induced
morphism of the cohomologies
\begin{eqnarray}\label{CohomMorph}
\varphi_{ab}^*:
H^*(\mathcal{F}_{a+b}(\mathcal{H}_{AB}))&\rightarrow&
H^*(\mathcal{F}_a(\mathcal{H}_A))\otimes
H^*(\mathcal{F}_b(\mathcal{H}_B))\nonumber\\
\sigma_w\qquad&\mapsto&\sum_{u,v} c^{uv}_w(a,b)\cdot
\sigma_u\otimes\sigma_v.
\end{eqnarray}
One can find the details of their calculation in
\cite{Klyachko2004}. Note that $c^{uv}_w(a,b)=1$ for identical
permutations $u,v,w$. Hence we get for free the following {\it
basic inequality\/}
\begin{equation}\label{basic}
\sum_i a_{i}\lambda^A_i+\sum_j b_{j}\lambda^B_j\le \sum_{k}
(a+b)^\downarrow_{k}\lambda^{AB}_k
\end{equation}
valid for all test spectra $a,b$.
\end{example}

%

\section{One point $\nu$-representability}
\label{H_lambda} In this section we apply the above results to
the morphism
$f:\mathrm{U}(\mathcal{H})\rightarrow\mathrm{U}(\mathcal{H}^\nu)$
given by an irreducible representation $\mathcal{H}^\nu$ of group
$\mathop{\mathrm{U}}(\mathcal{H})$ with a Young diagram $\nu$ of
order $N=|\nu|$. For a column diagram we return to the $N$-fermion
system $\wedge^N\mathcal{H}$, while a row diagram corresponds to
the $N$-boson space $S^N\mathcal{H}$. However, the main reason to
consider the general {\it para-statistical\/} representations
$\mathcal{H}^\nu$ is not a uniform treatment of fermions and
bosons, but  taking into account spin.
Observe that the state space of a single particle with spin splits
into the tensor product
$\mathcal{H}=\mathcal{H}_r\otimes\mathcal{H}_s$ of the orbital
$\mathcal{H}_r$ and the spin $\mathcal{H}_s$ degrees of freedom.
The total $N$-fermion space decomposes into spin-orbital
components as follows \cite{Weyl}
\begin{equation}\label{spin_orb}
\wedge^N(\mathcal{H}_r\otimes\mathcal{H}_s)=
\sum_{|\nu|=N}\mathcal{H}^\nu_r\otimes\mathcal{H}_s^{\nu^t},
\end{equation}
where $\nu^t$ stands for the transpose diagram. In many physical
systems, like electrons in an atom or a molecule, the total spin
is a well defined quantity that singles out a specific component
of this decomposition. Theorem \ref{Ber_Sja} applied to the
component gives all constraints on the possible spin and orbital
occupation numbers, see the details in
$n^\circ$~\ref{spin_orb_occ} below.

\subsection{Physical interpretation} Let's now relate Theorem 1 to
the $N$-representabi\-lity problem and its 
ramifications indicated above. We'll refer to the latter as the
$\nu$-{\it representability\/} problem.

It is instructive to think about $X\in\mathfrak{u}(\mathcal{H})$ as
an {\it observable\/} and
treat $\rho\in \mathfrak{u}(\mathcal{H})^*$ as a {\it mixed
state\/} with the duality pairing given by the expectation value
of $X$ in state $\rho$
\begin{equation}\label{expect}
\langle X,\rho\rangle=\mathop{\mathrm{Tr}_\mathcal{H}}X\rho
\end{equation}
(forget for a while about the positivity $\rho\ge0$ and
normalization $\mathop{\text{Tr}}\rho=1$).

We want to elucidate the physical meaning of the projection
$f^*:\mathfrak{u}(\mathcal{H}^\nu)^*\rightarrow
\mathfrak{u}(\mathcal{H})^*$ uniquely determined by the equation
$$\langle
f_*(X),\rho^\nu\rangle=\langle X,f^*(\rho^\nu)\rangle,\qquad
X\in\mathfrak{u}(\mathcal{H}),
\qquad\rho^\nu\in\mathfrak{u}(\mathcal{H}^\nu)^*.$$ In the above
setting (\ref{expect}) it reads as follows
\begin{equation}\label{proj1}
\mathop{\mathrm{Tr}_{\mathcal{H}^\nu}}(X\rho^\nu)=\mathop{\mathrm{Tr}_\mathcal{H}}(Xf^*(\rho^\nu)),\qquad\forall
X\in\mathfrak{u}(\mathcal{H}).
\end{equation}
A good point to start with is {\it Schur's duality\/} between
irreducible representations of the unitary
$\mathop{\mathrm{U}}(\mathcal{H})$ and the symmetric $S_N$ groups
\begin{equation}\label{schur}
\mathcal{H}^{\otimes
N}=\sum_{|\nu|=N}\mathcal{H}^\nu\otimes\mathcal{S}^\nu.
\end{equation}
The latter group acts on $\mathcal{H}^{\otimes N}$ by permutations
of the tensor factors, and its irreducible representations
$\mathcal{S}^\nu$ show up in the right hand side.
 One can treat
$\mathcal{H}^{\otimes N}$ as a state space of $N$-particles, and
for identical particles all physical quantities should commute with
$S_N$.
Looking into the right hand side of (\ref{schur}) we see that such
quantities are linear combinations of operators $\rho^\nu\otimes
1$ acting in the component $\mathcal{H}^\nu\otimes\mathcal{S}^\nu$
and equal to zero elsewhere. In the case of a genuine mixed state
$\rho^\nu$, i.e. a nonnegative operator of trace $1$, one can
treat $(\rho^\nu\otimes 1)/\dim \mathcal{S}^\nu $ as a mixed state
of $N$ identical particles obeying some para-statistics of type
$\nu$.
Let  $\rho_i:\mathcal{H}$ be its $i$-th {\it reduced state\/}.
Since $\rho^\nu\otimes 1$ commutes with $S_N$,
the reduced state $\rho=\rho_i$ is actually independent of $i$.
However, occasionally we retain the index $i$ just to indicate the
tensor component where it operates.
\begin{proposition}\label{claim} In the above notations
\begin{equation}\label{proj}
f^*(\rho^\nu)=N\rho.
\end{equation}
\end{proposition}
\proof We have to check that (\ref{proj}) fits the equation
(\ref{proj1}):
$$\mathop{\mathrm{Tr}_{\mathcal{H}^\nu}}(X\rho^\nu)=
\mathop{\mathrm{Tr}_{\mathcal{H}^\nu\otimes\mathcal{S}^\nu}}X\frac{\rho^\nu\otimes
1}{\dim\mathcal{S}^\nu}= \mathop{\mathrm{Tr}_{\mathcal{H}^{\otimes
N}}}X\frac{\rho^\nu\otimes 1}{\dim\mathcal{S}^\nu}=
\sum_i\mathop{\mathrm{Tr}_{\mathcal{H}}}X_i\rho_i=
N\mathop{\mathrm{Tr}_{\mathcal{H}}}X\rho,
$$
where $X_i$ is a copy of $X$ acting in the $i$-th component of
$\mathcal{H}^{\otimes N}$, so that
$$\mathop{\mathrm{Tr}_{\mathcal{H}^{\otimes
N}}}X_i\frac{\rho^\nu\otimes 1}{\dim\mathcal{S}^\nu}=
\mathop{\mathrm{Tr}_{\mathcal{H}}}X_i\rho_i$$ by definition
(\ref{reduced}) of reduced state. \qed

A general $\nu$-{\it representability problem\/} concerns with the
relationship between the spectrum $\mu$ of a mixed state
$\rho^\nu$ and spectrum $\lambda$ of its {\it particle density
matrix\/} $N\rho$.
The latter spectrum is known as the {\it occupation numbers\/}
\footnote{More precisely, the occupation numbers of {\it natural
orbitals\/}. The latter are defined as eigenvectors of the particle
density matrix.} of the system in state $\rho^\nu$. Formally the
constraints on the spectra are given by Theorem \ref{Ber_Sja}.

\begin{remark}
The above construction  allows for a given mixed state $\rho^\nu$
to define the higher order reduced matrices. Their
characterization would have almost unlimited applications. Indeed,
behavior of most systems of physical interest is governed by
two-particle interaction.
As  a result,  the energy of a state becomes  a linear functional
of its two-point reduced matrix. To minimize the energy and to
find the correlation matrix of the ground state one has to
elucidate all the constraints that a two-point reduced matrix
should satisfy. This problem and the whole program are known as
{\it Coulson challenge\/}\footnote{And also as two-particle
$N$-representability or, following D.~Herschbach,  as a  holy
grail of theoretical chemistry.} \cite{Coleman_Yukalov}. In the
form just described it may be unfeasible even for quantum
computers \cite{Christ07}. For other approaches and the current
state of art see
\cite{Mazziotti07}.
This problem is far beyond the scope of our paper. Nevertheless,
the characterization of one point reduced matrices given below
imposes also new constraints on the higher reduced states.
\end{remark}

\subsubsection{\it Constraints on spin and orbital occupation numbers}
\label{spin_orb_occ}
 Let's
return to a system of  $N$ fermions, this time of smallest
possible spin $s=1/2$, $\dim\mathcal{H}_s=2$. In this case
spin-orbital decomposition (\ref{spin_orb}) involves only terms
\begin{equation}\label{term}
\mathcal{H}_r^\nu\otimes\mathcal{H}_{s}^{\nu^t}
\end{equation}
with at most two-column diagram $\nu$. The sizes of the columns
$\alpha\ge \beta$ are determined by equations
\begin{equation}\label{ab}
\alpha+\beta=N,\qquad \alpha-\beta=2J,
\end{equation} where $J$ is the total spin of the system, so
that $\mathcal{H}_s^{\nu^t}=\mathcal{H}_J$ is just the spin $J$
representation of the group
$\mathop{\mathrm{SU}}(\mathcal{H}_s)=\mathop{\mathrm{SU}}(2)$.

Consider now a pure $N$-fermion state of total spin $J$
$$\psi\in \mathcal{H}_r^\nu\otimes\mathcal{H}_J,$$
where the diagram $\nu$ is determined by equations (\ref{ab}). Let
$\rho^\nu$ and $\rho^J$ be its reduced states in the orbital and
spin components respectively. The basic fact is that
the reduced states are isospectral
$\mathop{\text{Spec}}\rho^\nu=\mathop{\text{Spec}}\rho^J$.
Hence $\mathop{\text{Spec}}\rho^\nu$ can be identified with the
{\it spin occupation numbers\/}. On the other hand Theorem
\ref{Ber_Sja}, in view of Proposition \ref{claim}, relates
$\mathop{\text{Spec}}\rho^\nu$ with the {\it orbital occupation
numbers\/} given by the spectrum of the particle density matrix
$N\rho$. In this way one can produce all constraints on allowed
spin and orbital occupation numbers, {\it provided\/} that a
solution of the $\nu$-representability problem is known for
two-column diagrams. We address this issue  in sections
\ref{nu_rpr} and \ref{calc_coeff}. See also Corollary
\ref{spin_orb_ineq} in section \ref{nu_rpr}.





\subsection{Formal solution of the $\nu$-representability problem}
\label{nu_rpr}
Henceforth  we treat the lower index $r$ as the rank of the
Hilbert space $\mathcal{H}_r$. Recall that the character of the
representation $\mathcal{H}_r^\nu$, i.e. the trace of a diagonal
operator
\begin{equation}\label{diag}
z=\mbox{diag}(z_1,z_2,\ldots,z_r)\in\mathop{\mathrm{U}}(\mathcal{H}_r),
\end{equation} in some orthonormal basis $e$ of  $\mathcal{H}_r$, is
given by {\it Schur's function\/} $S_\nu(z_1,z_2,\ldots,z_r)$. It
has a purely combinatorial description in terms of the so called
{\it semistandard tableaux\/} $T$ of shape $\nu$. The latter are
obtained from the diagram $\nu$ by filling it with numbers
$1,2,\ldots, r$ strictly increasing  in columns and weakly in
rows.
Then the Schur function can be written as a sum of monomials
$z^T=\prod_{i\in T}z_i$
\begin{equation*}S_\nu(z)=\sum_T z^T
\end{equation*}
corresponding to all semistandard tableaux $T$ of shape $\nu$. The
monomials are actually the {\it weights\/} of representation
$\mathcal{H}_r^\nu$, meaning that
\begin{equation}\label{weight}
z\cdot e_T=z^Te_T\
\end{equation} for some basis $e_T$ of $\mathcal{H}^\nu_r$ parameterized by the
semistandard tableaux.
Denote by $\mathfrak{t}\subset\mathfrak{u}(\mathcal{H}_r)$ and
$\mathfrak{t}_\nu\subset\mathfrak{u}(\mathcal{H}_r^\nu)$ the
Cartan subalgebras of real diagonal operators in the bases $e$ and
$e_T$ respectively,
so that the {\it differential\/} of the above group action
$z:e_T\mapsto z^Te_T$ gives the morphism
\begin{equation}
f_*:\mathfrak{t}\rightarrow\mathfrak{t}_\nu,\qquad f_*(a):e_T\mapsto
a_Te_T,
\end{equation}
where $a_T:=\sum_{i\in T}a_i$. As in Example \ref{quant_mrg} we
treat the orbits $\mathcal{O}_a$ and $\mathcal{O}_{f_*(a)}$ as
flag varieties $\mathcal{F}_a(\mathcal{H}_r)$ and
$\mathcal{F}_{a^\nu}(\mathcal{H}_r^\nu)$ consisting of Hermitian
operators of spectra $a:a_1\ge a_2\ge \cdots\ge a_r$ and $a^\nu$
respectively. Here $a^\nu$ consists of the quantities $a_T$
arranged in the non-increasing order
\begin{equation}\label{a_nu}
a^\nu:=\{a_T\mid T=\text{semistandard tableau  of shape
}\nu\}^\downarrow.
\end{equation}
Finally, we need the morphism
\begin{equation}\label{phi}
\varphi_a:\mathcal{F}_a(\mathcal{H}_r)\rightarrow
\mathcal{F}_{a^\nu}(\mathcal{H}_r^\nu),\qquad X\mapsto f_*(X),
\end{equation}
together with its cohomological version
\begin{equation}\label{phi_coh}
\varphi_a^*:H^*(\mathcal{F}_{a^\nu}(\mathcal{H}_r^\nu))\rightarrow
H^*(\mathcal{F}_a(\mathcal{H}_r)),
\end{equation}
given in the canonical bases by coefficients $c_w^v(a)$:
\begin{equation}\label{c_w_v}
\varphi_a^*:\sigma_w\mapsto \sum_v c_w^v(a)\sigma_v.
\end{equation}
\begin{theorem}\label{nu thm} In the above notations all
constraints on the occupation numbers $\lambda$ of the system
$\mathcal{H}_r^\nu$ in a state $\rho^\nu$ of spectrum $\mu$ are
given by the inequalities
\begin{equation}\label{mix_nu_inq} \sum_i a_{i}\lambda_{v(i)}\le
\sum_k  a^\nu_{k}\mu_{w(k)}
\end{equation}
for all test spectra $a$ and permutations $v,w$ such that
$c_w^v(a)\neq 0$.
\end{theorem}
\proof In view of Proposition \ref{claim}, this is what Theorem
\ref{Ber_Sja} tells. One has to remember that the left action of a
permutation on ``places" is inverse to its right action on
indices. That is why the permutations $v$ and $w$, acting on $a$
and $f_*(a)=a^\nu$ in Theorem \ref{Ber_Sja}, move to the indices
of $\lambda$ and $\mu$ in the inequality (\ref{mix_nu_inq}). \qed

The coefficient $c_w^v(a)$ depends only on the order in which
quantities $a_T$ appear in the spectrum $a^\nu$. The order changes
when the test spectrum $a$ crosses a hyperplane
$$H_{T|T^\prime}: \sum_{i\in T}a_i=\sum_{j\in T^\prime}a_j.$$
The hyperplanes  cut the set of all test  spectra into a finite
number of
polyhedral cones called {\it cubicles\/}. For each cubicle one has
to check the inequality (\ref{mix_nu_inq}) only for its {\em
extremal edges\/}. As a result, the $\nu$-representability amounts
to a {\em finite system} of linear inequalities.

\begin{remark}
Let's emphasize once again the  difference between
Berenstein-Sjamaar theorem \cite[Thm 3.2.1]{Ber-Sja} and its
version
used in this paper. In the settings of Theorem~\ref{nu thm} it
manifests itself in the way    how the quantities $a_T$ are
ordered in the spectrum $a^\nu$, or what is the same which
parabolic subgroup is used for definition of Schubert cocycles.
Berenstein and Sjamaar choose a specific order of tableaux $T$,
while we rely on the natural order of the quantities
$a_T=\sum_{i\in T}a_i$. The latter choice allows to treat the
inequalities uniformly, and to avoid a rather cumbersome
transformation every time the test spectrum passes from one
cubicle to another.

Recall from $n^\circ$~\ref{spin_orb_occ} that the theorem also
describes  a relationship  between the spin and orbital occupation
numbers. We keep for them the above notations $\mu$ and $\lambda$
respectively.

\begin{corollary}\label{spin_orb_ineq} All constraints on spin
 and orbital  occupation numbers of $N$-electron
system in a pure state of total spin $J$ are given by the
inequalities {\rm (\ref{mix_nu_inq})}, applied to two column
diagram $\nu$ determined by equations {\rm (\ref{ab})}, and
bounded to mixed states $\rho^\nu$ of rank not exceeding
dimensionality $2J+1$ of the spin space.\qed
\end{corollary}

We postpone the calculation of the coefficients $c_w^v(a)$ to 
section~\ref{calc_coeff} and
focus instead
on some
general results that can be deduced from the
theorem as it stands.

%
%
\end{remark}
\subsubsection {\it Basic inequalities\/}
Being a ring homomorphism, $\varphi^*_a$ maps unit into unit
$\varphi^*_a(1)=1$, that is $c_w^v(a)=1$ for identical
permutations $v,w$. Hence the following {\it basic inequality\/}
$$\sum_i a_i\lambda_i\le
\sum_k  a^\nu_k\mu_k$$ holds for all test spectra $a$. Let's look at
it more closely for a {\it pure state\/}
$\rho^\nu=|\psi\rangle\langle\psi|$ in which case the right hand
side is maximal
and the inequality takes the form
\begin{equation}\label{basic_pure}
\sum_i a_i\lambda_i\le a^\nu_1=\max_T
\sum_{i\in T}a_i=\sum_i a_i\nu_i,
\end{equation} where
$\nu_1\ge\nu_2\ge\cdots\ge 0$ are rows of $\nu$. The maximum in
the right hand side is attained for the tableau $T$ of shape $\nu$
whose $i$-row is filled by $i$.

The normalization $\sum_i\lambda_i=N=\sum_j\nu_j$ allows to shift
the test spectra into the positive domain $a_1\ge a_2\ge
\cdots\ge 0$, so that they became nonnegative linear combinations of the
fundamental weights
\begin{equation}\label{fund_wght}
\omega_k=(\underbrace{1,1,\ldots,1}_k,0,0,\ldots,0).
\end{equation} 
Hence it is enough to check (\ref{basic_pure}) for $a=\omega_k$,
that gives the {\it majorization inequality\/}
$\lambda\preceq\nu$, cf. Example \ref{U(n)}. Thus we arrived
at the first claim of the following result that characterizes
occupation numbers of system $\mathcal{H}^\nu$ in an unspecified
mixed state.

\begin{theorem}\label{mixed_occ}
The occupation numbers of the system $\mathcal{H}^\nu$ in an
arbitrary mixed state satisfy the {\it majorization inequality\/}
\begin{equation}\label{maj}
\lambda\preceq
\nu,
\end{equation}
and any such $\lambda$ can be realized as the occupation numbers
of
some mixed state.
\end{theorem}
\proof 
The second claim follows from two observations:
\begin{enumerate}
\item {\it The occupation numbers of
a coherent state $\psi\in\mathcal{H}^\nu$, that is a highest
vector of the representation, are equal to $\nu$.\/}
\item {\it The
set of allowed occupation numbers, written in any order, form a
convex set.\/}
\end{enumerate}
Indeed, the polytope given by the majorization inequality
(\ref{maj}) is just a convex hull of vectors obtained from $\nu$
by permutations of coordinates, cf. Example \ref{U(n)}. Hence by 1
and 2 it consists of legitimate occupation numbers.

{\it Proof of 1.\/} Consider a decomposition of the complexified
Lie algebra
$$\mathfrak{u}(\mathcal{H})\otimes\mathbb{C}=\mathfrak{gl}(\mathcal{H})=
\mathfrak{n}_-+\mathfrak{h}+\mathfrak{n}_+,
$$
into a diagonal Cartan subalgebra
$\mathfrak{h}=\mathfrak{t}\otimes
\mathbb{C}$ accompanied with lower- and upper-triangular nilpotent
subalgebras $\mathfrak{n}_\mp$. By definition $\mathfrak{n}_+$
annihilates the highest vector $\psi\in\mathcal{H}^\nu$ of weight
$\nu$. Hence $\langle\psi|X^\pm|\psi\rangle=\langle X^\mp
\psi|\psi\rangle=0$ for all $X^\pm\in\mathfrak{n}_\pm$. Then by
equation (\ref{proj1})
$$\langle\psi|X^\pm|\psi\rangle=
\mathop{\mathrm{Tr}_{\mathcal{H}^\nu}}(X^\pm|\psi\rangle\langle\psi|)=
\mathop{\mathrm{Tr}_\mathcal{H}}(X^\pm f^*(|\psi\rangle\langle\psi|))=0,\qquad
\forall X^\pm\in\mathfrak{n}_\pm.$$
This means that $\rho=f^*(|\psi\rangle\langle\psi|)$ is a diagonal
matrix. On the other hand $t\psi=\langle t,\nu\rangle\psi$ for
$t\in
\mathfrak{t}$, hence as above
$$\langle t,\nu\rangle=\langle\psi|t|\psi\rangle=
\mathop{\mathrm{Tr}_{\mathcal{H}^\nu}}(t|\psi\rangle\langle\psi|)=
\mathop{\mathrm{Tr}_\mathcal{H}}(t f^*(|\psi\rangle\langle\psi|))=
\mathop{\mathrm{Tr}_\mathcal{H}}(t\rho)=\langle t,\rho\rangle,$$
that is $\mathop{\mathrm{Spec}}\rho=\nu$.

{\it Proof of 2.\/} Let  $\rho^\nu_1$, $\rho^\nu_2$ be mixed
states, with the particle densities $\rho_1$, $\rho_2$, and the
occupation numbers $\lambda_1$, $\lambda_2$. Apply to $\rho_1$,
$\rho_1^\nu$ a unitary rotation $\rho_1\mapsto U\rho_1U^*$,
$\rho_1^\nu\mapsto U\rho_1^\nu U^*$ that transforms orthonormal
eigenvectors of $\rho_1$ into that of $\rho_2$ in a prescribed
order.
The resulting new operators $\rho_1$, $\rho_2$ commute and have
the original spectra $\lambda_1,\lambda_2$. Then the particle
density matrix $\rho=p_1\rho_1+p_2\rho_2$ of the convex
combination $\rho^\nu=p_1\rho_1^\nu+p_2\rho_2^\nu$ has spectrum
$\lambda=p_1\lambda_1+p_2\lambda_2$. \qed

For a column diagram $\nu$ the majorization inequality
$\lambda\preceq\nu$ amounts to the {\it Pauli exclusion
principle\/} $\lambda_i\le 1$. In general, we refer to it as the
{\it Pauli constraint\/}.
Note that the above proof shows that equality in (\ref{maj}) is
attained for the coherent states only.
The second part of Theorem \ref{mixed_occ} extends Coleman's
result \cite{Coleman63} for $\wedge^N\mathcal{H}$.

Recall, that the theorem solves the $\nu$-representability problem
for
unspecified mixed states. We will see later that for pure states
the answer in general is much more complicate. Nevertheless, there
are surprisingly many systems for which the majorization
inequality along is sufficient for pure $\nu$-representability. We
address them in the next item.


\subsubsection {\it Pure moment polytope\/}\label{moment}
One of the most striking features of Theorem \ref{nu thm} is the
linearity of the constraints (\ref{mix_nu_inq}). As a result, the
allowed spectra $(\lambda,\mu)$ form a convex polytope, called
(noncommutative) {\it moment polytope\/}. The convexity still
holds for any fixed $\mu=\mathop{\text{Spec}}\rho^\nu$, and in
particular for the occupation numbers $\lambda$ of all {\it
pure\/} states. We refer to the latter case as the {\it pure
moment polytope\/}. It sits inside the positive Weyl chamber, and
its multiple kaleidoscopic reflections in the walls of the chamber
generally form a {\it nonconvex\/} rosette, consisting of all
legitimate occupation numbers written in an arbitrary order. It
can be convex only if all constraints on the occupation numbers
are given by the majorization inequality $\lambda\preceq\nu$
alone. Here we describe a class of representations
$\mathcal{H}^\nu$ with this property.

This happens, for example,
for a system of $N\ge2$ bosons. In this case $\nu$ is a row
diagram and the majorization inequality imposes no
constraints on 
$\lambda$. By Theorem
\ref{mixed_occ} this means that every nonnegative spectrum $\lambda$
of trace $N$ represents occupation numbers of some {\it mixed\/}
state.
However for bosons one can easily find a {\it pure\/} state that
does the job:
$$\psi=\sum_{i}
\sqrt{\lambda_i}e_i^{\otimes N}\in S^N\mathcal{H},$$
where $e_i$ is an orthonormal basis of $\mathcal{H}$.
 This
makes the bosonic $N$-representa\-bi\-lity problem meaningless.


A more interesting physical  example constitutes the so-called
{\it closed shell\/}, meaning a system of electrons of total spin
zero. The corresponding diagram $\nu$ consists of two columns of
equal length. We will see shortly that in this case the Pauli
constraint $\lambda\le 2$ shapes the pure moment polytope.

Observe that it is enough to construct pure states whose
occupation numbers are generators of the cone
cut out of the Weyl chamber by the majorization inequality
$\lambda\preceq\nu$. Then the convexity does the rest.

Recall, that in the proof of Theorem \ref{mixed_occ} we have
already identified
$\nu$ with the occupation numbers of a {\it coherent state\/}.
Due to the majorization inequality $\lambda\preceq\nu$, the
entropy of its reduced state is minimal possible.
By that reason coherent states are
generally considered as closest to classical ones
\cite{Perelomov}.
At the other extreme one finds the so called {\it completely
entangled\/} states $\psi\in\mathcal{H}^\nu$ whose particle
density matrix $\rho=f^*(|\psi\rangle\langle\psi|)$ is scalar and
the reduced entropy is maximal \cite{Kl07}. By definition
(\ref{proj1}) we have
$\mathop{\text{Tr}_\mathcal{H}}(X\rho)=\mathop{\text{Tr}_{\mathcal{H}^\nu}}(X|\psi\rangle\langle\psi|)
=\langle\psi|X|\psi\rangle$, so that the completely entangled
states can be  described by equation
\begin{equation}\label{ent_eqn}
\langle\psi|X|\psi\rangle=0,\qquad\forall\;
X\in\mathfrak{su}(\mathcal{H}).
\end{equation}
Let's call a system $\mathcal{H}_r^\nu$ {\it exceptional\/} if the
$\mathop{\mathrm{SU}}(\mathcal{H}_r)$-representation
$\mathcal{H}_r^\nu$ is equivalent to one of the following:
$\mathcal{H}_r$, its dual $\mathcal{H}^*_r$, and, for odd rank
$r$, $\wedge^2\mathcal{H}_r$, $\wedge^2\mathcal{H}^*_r$.  The
Young diagram $\nu$ of an exceptional system can be obtained from
$r\times m$ rectangle  by adding an extra column of length
$1,r-1,2,r-2$ respectively.

One
readily realizes  that the exceptional systems contain no
completely entangled states, say because reduced matrix of
 $\psi\in\wedge^2\mathcal{H}_r$ has an even rank.
\begin{proposition}\label{non_excep}
In every non-exceptional system $\mathcal{H}^\nu$ there exists a
completely entangled state.
\end{proposition}
\proof The result is actually well known, but in a different
context. 
The entanglement equation (\ref{ent_eqn}) is nothing but the
stationarity condition  for the length of vector
$\langle\psi|\psi\rangle$ with respect to action of the {\it
complexified\/} group $\mathop{\mathrm{SL}}(\mathcal{H})$. It is
known \cite{Vinberg} that every stationary point is actually a
minimum, and an $\mathop{\mathrm{SL}}(\mathcal{H})$-orbit contains
a minimal vector if and only if the orbit is closed.
As a result, we end up
with the problem of existence of a nonzero closed orbit, or, what
is the same, the existence of a nonconstant polynomial invariant.
The proposition just reproduces a known answer to the latter
question \cite{Vinberg}. \qed

By admitting  other simple Lie groups we find only two more
exceptional representations: the standard representation of the
symplectic group $\mathop{\mathrm{Sp}}(n)$ and a halfspinor
representation of $\mathop{\mathrm{Spin}}(10)$.

Now we
can solve the pure $\nu$-representability problem for a wide class
of systems, including the above mentioned closed shell.

\begin{theorem}\label{mlt_clmn}
Suppose that all columns of Young diagram $\nu$ are multiple,
meaning that every number in the sequence of columns lengths
$\nu_1^t\ge\nu_2^t\ge \nu_3^t\ge\cdots$ appears at least twice.
Then all constraints on the occupation numbers of the system
$\mathcal{H}^\nu$ in a pure state are given by the majorization
inequality $\lambda\preceq \nu$ along.
\end{theorem}
\proof
We'll proceed by induction on the height of the diagram $\nu$. The
triviality of the bosonic $N$-representability problem provides a
starting point for the induction.

Let now $\lambda$ be a vertex of the polytope cut out of the
positive Weyl chamber by the majorization inequality
$\lambda\preceq\nu$. Take notice that the latter  includes
equation $\mathop{\mathrm{Tr}}\lambda=\mathop{\mathrm{Tr}}\nu$.
Then the following alternative holds:
\begin{enumerate}
\item {\it Either all nonzero components of $\lambda$ are equal,}
\item {\it Or one can
split $\lambda$ and $\nu$ into two parts
$\lambda=\lambda^\prime|\lambda^{\prime\prime}$,
$\nu=\nu^\prime|\nu^{\prime\prime}$ containing the first $p$
components and  the remaining ones, both satisfying the
inequalities $\lambda^\prime\preceq\nu^\prime$,
$\lambda^{\prime\prime}\preceq\nu^{\prime\prime}$.}
\end{enumerate}
Indeed, the second claim just tells that the $p$-th majorization
inequality in (\ref{maj_inq}) turns into equation. On the other
hand, if all the majorization inequalities are strict, and
$\lambda$ contains different nonzero entries, then one can
linearly vary these entries preserving the non-increasing order of
$\lambda$ and the majorization $\lambda\preceq\nu$. As result we
get a line segment in the polytope containing $\lambda$,
which is impossible for a vertex.

We've to prove that every vertex $\lambda$ represents occupation
numbers of some pure state. Consider the above two cases
separately.

\begin{case} 
Let $\lambda$ contains $r$ equal nonzero entries and
$\mathcal{H}_r\subset\mathcal{H}$ be a subspace of dimension $r$.
The conditions of the theorem ensure that the system
$\mathcal{H}_r^\nu$ is non-exceptional, hence by Proposition
\ref{non_excep} it contains a state $\psi\in\mathcal{H}^\nu_r$ with
occupation numbers equal to nonzero part of $\lambda$.  In bigger
system $\mathcal{H}^\nu\supset\mathcal{H}^\nu_r$ its occupation
numbers will be extended by zeros.
\end{case}

\begin{case} Let the system has rank $r=p+q$. Choose a
decomposition $\mathcal{H}_r=\mathcal{H}_p\oplus \mathcal{H}_q$
and consider a restriction of the representation
$\mathcal{H}_r^\nu$ onto subgroup
$\mathrm{U}(\mathcal{H}_p)\times\mathrm{U}(\mathcal{H}_q)$
\begin{equation}\label{L-R}
\mathcal{H}_r^\nu=\sum_{\mu,\pi} c_{\mu\pi}^\nu\mathcal{H}_p^\mu\otimes
\mathcal{H}_q^\pi,
\end{equation}
where $c_{\mu\pi}^\nu$ are the omnipresent Littlewood-Richardson
coefficients. Observe that $c_{\nu^\prime\nu^{\prime\prime}}^\nu=1$,
and therefore $\mathcal{H}_p^{\nu^\prime}\otimes
\mathcal{H}_q^{\nu^{\prime\prime}}\subset\mathcal{H}_r^\nu$. By
induction hypothesis there exist states
$\psi^\prime\in\mathcal{H}_p^{\nu^\prime}$ and
$\psi^{\prime\prime}\in\mathcal{H}_q^{\nu^{\prime\prime}}$ with
occupation numbers $\lambda^\prime$, $\lambda^{\prime\prime}$ and
particle densities  $\rho^\prime$, $\rho^{\prime\prime}$
respectively. Then decomposable state
$\psi=\psi^\prime\otimes\psi^{\prime\prime}$ has particle density
$\rho^\prime\oplus\rho^{\prime\prime}$, and its occupation numbers
are equal to $\lambda=\lambda^\prime|\lambda^{\prime\prime}$. \qed
\end{case}
Let's extract for a reference a useful corollary from the last
part of the proof.
\begin{corollary}\label{cor_mlt_clmn} Suppose that the Littlewood-Richardson coefficient
$c_{\mu\pi}^\nu$ is nonzero. Then merging of the occupation
numbers $\lambda^\prime$, $\lambda^{\prime\prime}$ of the systems
$\mathcal{H}_p^\mu$, $\mathcal{H}_q^\pi$ form legitimate
occupation numbers of the system $\mathcal{H}^\nu_{p+q}$. \qed
\end{corollary}

\begin{remark}\label{extra_syst}
The restriction on the column's multiplicities of the diagram
$\nu$ is needed only to ensure that the components of any
splitting
$\nu=\nu^\prime|\nu^{\prime\prime}|\nu^{\prime\prime\prime}|\ldots$
are non-exceptional. The latter condition holds for any two-row
diagram $[\alpha,\beta], \beta\ne1$  for $\dim\mathcal{H}\ge 3$.
This gives examples of systems  beyond Theorem~\ref{mlt_clmn}, say
for $\nu=[3,2]$,
whose pure moment polytope is given by the majorization inequality
along. More such diagrams can be produced as follows: take  $\nu$
as in Theorem~\ref{mlt_clmn} and remove one cell from its last
row. This works when the last row contains at least three cells
and rank of the system is bigger than the height
of $\nu$. 
A complete classification of all such systems is still missing.
\end{remark}
\subsubsection{Dadok-Kac construction}\label{Dad_Kac}

In the last two theorems we encounter the problem of construction
a pure state with given occupation numbers. The problem lies at
the very heart of the $\nu$-representability and one shouldn't
expect an easy solution. Nevertheless, there is a combinatorial
construction that
produces a state with {\it diagonal\/} density matrix, whose
spectrum can be easily controlled. It has been used first by
Borland and Dennis \cite{Borland Dennis} to forecast the structure
of the moment polytope for small fermionic systems. Later on
M\"uller \cite{Muller} formalized and advanced
their approach to the limit. It fits into a general Dadok-Kac
construction \cite{Dadok-Kac} that works for any representation.

Below we follow the notations introduced at the beginning of
$n^\circ$~\ref{nu_rpr}.
Let $x=\mathrm{diag}(x_1,x_2,\ldots,x_r)$ be a typical element
from Cartan subalgebra
$\frak{t}\subset\mathfrak{u}(\mathcal{H}_r)$. For a given
semi-standard tableau $T$ call the linear form $\omega_T:x\mapsto
x_T=\sum_{i\in T}x_i$ the {\it weight\/} of the basic vector
$e_T\in
\mathcal{H}^\nu_r$.
We also need nonzero weights of the adjoint representation
$\alpha_{ij}:x\mapsto x_i-x_j$, $i\ne j$
called {\it roots\/}. Let's turn the set of 
semi-standard tableaux of shape $\nu$ into a graph by
connecting $T$ and $T^\prime$ each time $\omega_T-\omega_{T^\prime}$
is a root, i.e. the contents of $T$ and $T^\prime$, considered as
multi-sets, differ by exactly one element.

\begin{proposition}\label{Dad-Kac} Let $\mathbf{T}$ be a set of
semi-standard tableaux of shape $\nu$ containing no connected pairs.
Then every state
$\psi=\sum_{T\in\mathbf{T}}c_Te_T\in\mathcal{H}^\nu$ with support
$\mathbf{T}$ has a diagonal particle density matrix with entries
\begin{equation}\label{red_diag}
\lambda_i=\sum_{T\ni i}|c_T|^2,
\end{equation}
where every tableau $T$ is counted as many times as the index $i$
appears in it.
\end{proposition}
\proof The proof refines the arguments used in claim 1 of Theorem
\ref{mixed_occ}, from which we borrow the notations. As in the
above theorem we have to prove $\langle\psi|X|\psi\rangle=0$ for
every $X\in\mathfrak{n}_++\mathfrak{n}_-$. It is enough to
consider root vectors $X_\alpha$ that form a basis of
$\mathfrak{n}_++\mathfrak{n}_-$. Then
$$\langle\psi|X_\alpha|\psi\rangle=
\sum_{T,T^\prime\in\mathbf{T}}\overline{c}_{T^\prime}c_T\langle
e_{T^\prime}|X_\alpha|e_T\rangle.$$ Since  $X_\alpha e_T$ has weight
$\alpha+\omega_T$, it is orthogonal to $e_{T^\prime}$, except
$\omega_{T^\prime}=\omega_T+\alpha$. The latter is impossible for
$T,T^\prime\in\mathbf{T}$, and therefore the reduced state of $\psi$
is diagonal. A straightforward calculation gives the diagonal
entries (\ref{red_diag}). \qed

We'll have a chance to use this construction in
$n^\circ$~\ref{2row}.

 Take notice
that for a fixed support $\mathbf{T}$ the set of unordered spectra
(\ref{red_diag}) form a convex polytope. It is not known when this
approach exhausts the whole moment polytope. The smallest
fermionic system where it fails is $\wedge^3\mathcal{H}_8$, see
$n^\circ$~\ref{small}.

\subsection{Calculation of the coefficients $c_w^v(a)$} \label{calc_coeff}
To move further and to give Theorem \ref{Ber_Sja} the full
strength one has to calculate the coefficients $c_w^v(a)$.
Berenstein and Sjamaar left this problem mostly untouched.
However, in the $\nu$-representability settings, highlighted in
Theorem \ref{nu thm}, this can be done pretty explicitly.

\subsubsection{Canonical generators} To proceed  we first need an alternative  description
of the cohomology of flag variety
$\mathcal{F}_a(\mathcal{H}_r)$ \cite{BGG}. Recall that the latter
understood here as the set of Hermitian operators in
$\mathcal{H}_r$ of given spectrum $a$. To avoid technicalities, we
assume the spectrum to be
simple $a_1>a_2>\cdots>a_r$. Let $\mathcal{E}_i$ be the {\it
eigenbundle\/} on $\mathcal{F}_a(\mathcal{H}_r)$ whose fiber at
$X\in\mathcal{F}_a(\mathcal{H}_r)$ is the eigenspace of operator
$X$ with eigenvalue $a_i$.  Their Chern classes
$x_i=c_1(\mathcal{E}_i)$ generate the cohomology ring
$H^*(\mathcal{F}_a(\mathcal{H}_r))$ and we refer to them as the
{\it canonical generators\/}. The elementary symmetric functions
$\sigma_i(x)$ of the canonical generators are the characteristic
classes of the trivial bundle $\mathcal{H}_r$ and thus vanish.
This identifies the cohomology with the {\it ring of
coinvariants\,}
\begin{equation}\label{canonical}
H^*(\mathcal{F}_a(\mathcal{H}_r))=\mathbb{Z}[x_1,x_2,\ldots,x_r]/(\sigma_1,\sigma_2,\ldots,\sigma_r).
\end{equation}
This approach to the cohomology is more functorial and by that
reason leads to an easy calculation of the morphism
(\ref{phi_coh})
$$\varphi_a^*:H^*(\mathcal{F}_{a^\nu}(\mathcal{H}^\nu))\rightarrow
H^*(\mathcal{F}_{a}(\mathcal{H})).$$
Recall that the spectrum $a^\nu$ consists of the quantities
$a_T=\sum_{i\in T}a_i$ arranged in decreasing order, where $T$
runs over all semi-standard tableaux of shape $\nu$. We define
$x_T=\sum_{i\in T}x_i$ in a similar way.
\begin{proposition}\label{prop_coh_mrf} Let $x_i$ and $x^\nu_k$ be the canonical
generators of $H^*(\mathcal{F}_{a}(\mathcal{H}))$ and
$H^*(\mathcal{F}_{a^\nu}(\mathcal{H}^\nu))$ respectively. Then
\begin{equation}\label{coh_mrf}
\varphi^*_a(x^\nu_k)= x_T,\quad\text{ when }\quad a^\nu_k= a_T.
\end{equation}
In other words, $\varphi^*_a(x^\nu_k)$ is obtained from $a_k^\nu$
by the substitution ${a_i\mapsto x_i}$.
\end{proposition}
\proof The eigenbundle $\mathcal{E}_i$ is equivariant with respect
to the adjoint action $X\mapsto uXu^*$ of the unitary group
$\mathop{\mathrm{U}}(\mathcal{H})$.
Therefore it is uniquely determined by the linear representation
of the centralizer $D=Z(X)$ in a fixed fiber $\mathcal{E}_i(X)$ or
by its character $\varepsilon_i:D\rightarrow \mathbb{S}^1=\{z\in
\mathbb{C}^*\mid |z|=1\}$. In the eigenbasis $e$ of the operator $X$ the
centralizer becomes a diagonal torus with typical element
$z=\text{diag}(z_1,z_2,\ldots,z_r)$ and the character
$\varepsilon_i:z\mapsto z_i$.

Let now $X^\nu=\varphi_a(X)$, $D^\nu=Z(X^\nu)$, and $e_T$ be the
weight basis of $\mathcal{H}^\nu$, introduced in section
\ref{nu_rpr}, parameterized by semi-standard tableaux $T$ of shape
$\nu$ and arranged in the order of eigenvalues $a^\nu$. Then the
character of the pull back $\varphi_a^{-1}(\mathcal{E}^\nu_k)$ is
just the weight $\prod_{i\in T}\varepsilon_i$ of the $k$-th vector
$e_T$, where the tableau $T$ is determined from the equation
$a^\nu_k=a_T$, cf. (\ref{weight}). Thus
$\varphi_a^{-1}(\mathcal{E}^\nu_k)=\bigotimes_{i\in
T}\mathcal{E}_i$ and
we finally get
$$\varphi_a^*(x^\nu_k)=\varphi_a^*(c_1(\mathcal{E}^\nu_k))=
c_1(\varphi_a^{-1}(\mathcal{E}^\nu_k))= c_1(\bigotimes_{i\in
T}\mathcal{E}_i)= \sum_{i\in T}x_i=x_T. \quad\qed$$


\begin{remark}\label{indep} Formula (\ref{coh_mrf}) may look ambiguous for
a degenerate spectrum $a$, while in fact it is perfectly
self-consistent.
Indeed, consider a small perturbation $\tilde{a}$, resolving
multiple components of $a$, and the natural projection
$$\pi:\mathcal{F}_{\tilde{a}}(\mathcal{H})\rightarrow\mathcal{F}_a(\mathcal{H})$$
that maps $\widetilde{X}=\sum_i\tilde{a}_i|e_i\rangle\langle e_i|$
into $X=\sum_i a_i|e_i\rangle\langle e_i|$, where $e_i$ is an
orthonormal eigenbasis of $\widetilde{X}$.
It is known \cite{BGG} that $\pi$ induces isomorphism 
\begin{equation}\label{deg_coh}
\pi^*:H^*(\mathcal{F}_a(\mathcal{H}))\simeq
H^*(\mathcal{F}_{\tilde{a}}(\mathcal{H}))^{W(D)},
\end{equation}
where on the right hand side stands algebra of invariants with
respect to permutations of the
canonical generators $\tilde{x}_i$ with the same unperturbed
eigenvalue $a_i=\alpha$. Such permutations form Weyl group $W(D)$
of the maximal torus $\widetilde{D}=Z(\widetilde{X})$ in $D=Z(X)$.
For example, characteristic classes of the eigenbundle
$\mathcal{E}_\alpha$ with multiple eigenvalue $\alpha=a_i$
correspond to elementary symmetric functions of the respective
variables $\tilde{x}_i$.

Equation (\ref{coh_mrf}), as it stands, depends on a specific
ordering of the unresolved spectral values $a_i$ and $a_k^\nu$.
However,
when $\varphi_a^*$ applied to 
invariant elements with respect to the above Weyl group, 
the ambiguity vanishes.

Note also, that Schubert cocycle $\sigma_w\in
H^*(\mathcal{F}_{\tilde{a}}(\mathcal{H}))$ is invariant with respect
to $W(D)$ if and only if $w$ is the shortest representative in its
left coset modulo $W(D)$. Such cocycles form the canonical basis of
cohomology $H^*(\mathcal{F}_a(\mathcal{H}))$.

%
%
%
%
%
%
\end{remark}

\subsubsection{Schubert polynomials}
To calculate the coefficients $c_w^v(a)$ we have to return back to
the Schubert cocycles $\sigma_w$ and express them via the
canonical generators $x_i$.
This can be accomplished by  the {\it divided difference
operators\,}
\begin{equation}
\partial_i:f(x_1,x_2,\ldots,x_n)\mapsto
\frac{f(\ldots,x_i,x_{i+1},\ldots)-f(\ldots,x_{i+1},x_i,\ldots)}
{x_i-x_{i+1}}
\end{equation}
as follows. Write a permutation $w\in S_n$ as a product of the
minimal number of transpositions $s_i=(i,i+1)$
\begin{equation}\label{decomp}
w=s_{i_1}s_{i_2}\cdots s_{i_\ell}.
\end{equation}
The number of factors $\ell(w)=\#\{i<j\mid w(i)>w(j)\}$ is called
the {\it length\,} of the permutation $w$.
The product
\begin{equation*}
\partial_w:=\partial_{i_1}\partial_{i_2}\cdots\partial_{i_\ell}
\end{equation*}
is independent of the reduced decomposition
and  in terms of these operators the Schubert cocycle $\sigma_w$
is given by the equation
\begin{equation}\label{Schub}
\sigma_w=\partial_{w^{-1}w_0}(x_1^{n-1}x_2^{n-2}\cdots x_{n-1}),
\end{equation}
where  $w_0=(n,n-1,\ldots,2,1)$ is the unique permutation of the
maximal length.

The right hand side of equation (\ref{Schub}) makes sense for
independent variables $x_i$ and in this setting it is called {\it
Schubert polynomial\/} $S_w(x_1,x_2,\ldots,x_n)$, $\deg
S_w=\ell(w)$. They where first introduced by Lascoux and
Sch\"utzen\-berger \cite{L-Sch,L-Sch82} who studied them in a long
series of papers. See \cite{Macdonald91} for further references
and a concise exposition of the theory. We borrow  from
\cite{L-Sch} the
following table, in which $x,y,z$ stand for $x_1,x_2,x_3$. 
\begin{center}{\scriptsize\label{SchPol}
\begin{tabular}[b]{|c|c||c|c||c|c||c|c|}
\hline
$w$ & $S_w$&$w$&$S_w$&$w$&$S_w$&$w$&$S_w$\\
\hline%
3210&{\scriptsize $x^3y^2z$}&2301&{\scriptsize $x^2y^2$}&
2031&{\scriptsize $x^2y+x^2z$}&1203&{\scriptsize $xy$}\\
\hline 2310&{\scriptsize $x^2y^2z$}&3021&{\scriptsize $x^3y+x^3z$}&
2103&{\scriptsize $x^2y$}&2013&{\scriptsize $x^2$}\\
\hline 3120&{\scriptsize $x^3yz$}&3102&{\scriptsize $x^3y$}&
3012&{\scriptsize $x^3$}&0132&{\scriptsize $x+y+z$}\\
\hline 3201&{\scriptsize $x^3y^2$}&1230&{\scriptsize $xyz$}&
0231&{\scriptsize $xy+yz+zx$}&0213&{\scriptsize $x+y$}\\
\hline 1320&{\scriptsize $x^2yz+xy^2z$}&0321&{\scriptsize
$x^2y+x^2z+xy^2$}&
0312&{\scriptsize $x^2+xy+y^2$}&1023&{\scriptsize $x$}\\
\hline 2130&{\scriptsize $x^2yz$}&1302&{\scriptsize $x^2y+xy^2$}&
1032&{\scriptsize $x^2+xy+xz$}&0123&{\scriptsize $1$}\\
\hline
\end{tabular}}
\end{center}

Extra variables $x_{n+1},x_{n+2},\ldots$ being added to
(\ref{Schub}) leave Schubert polynomials unaltered.
By that reason they are usually treated as
polynomials in an infinite ordered alphabet  $x=(x_1,x_2,\ldots)$.
With this understanding every homogeneous polynomial can be
decomposed into Schubert components
as follows
\begin{equation*}
f(x)=\sum_{\ell(w)=\deg(f)} \partial_w f\cdot S_w(x).
\end{equation*}
Applying this to the polynomial
\begin{equation*}
\varphi^*_a(S_w(x^\nu))=
S_w(\varphi^*_a(x^\nu))=\sum_{\ell(v)=\ell(w)} c^v_w(a)\cdot S_v(x),
\end{equation*}
and using Proposition \ref{prop_coh_mrf} we finally arrive at the
following result.
\begin{theorem}\label{nu_coeff} For the $\nu$-representability problem the
coefficients of the decomposition $\varphi_a^*(\sigma_w)=\sum_v
c_w^v(a)\sigma_v$ are given by the formula
\begin{equation}\label{SchCoeff}
c^v_w(a)=\partial_v S_w(x^\nu)\mid_{x^\nu_k\mapsto x_T},
\end{equation}
where the tableau $T$ is derived from equation $a^\nu_k=a_T$, and
the operator $\partial_v$ acts on the variables $x_i$, replacing
$x^\nu_k$ via specialization $x^\nu_k\mapsto x_T=\sum_{i\in
T}x_i$.
\qed
\end{theorem}
Take notice that this equation is independent of an ordering of
the unresolved spectral values $a_k^\nu$. Indeed, Schubert
polynomial $S_w(x^\nu)$ is symmetric in the respective variables
$x_k^\nu$, {\it provided\/} that $w$ is the minimal representative
in its left coset modulo centralizer of the spectrum $a^\nu$ in
the symmetric group. Only such permutations correspond to Schubert
cocycles $\sigma_w\in H^*(\mathcal{F}_{a^\nu}(\mathcal{H}^\nu))$,
cf. Remark~\ref{indep}.

%


\section{Beyond the basic constraints}\label{beyond}

Here we use the above results to derive some general inequalities
for the pure $\nu$-representability problem beyond the Pauli
constraint $\lambda\preceq\nu$. We start with a complete solution
of the
problem for two-row diagrams, and then turn to the initial
$N$-representability problem that appears to be the most difficult
one.


\subsection{Two-row diagrams}\label{2row}
For two-row diagram $\nu=[\alpha,\beta]$ the majorization
inequality $\lambda\preceq\nu$ just tells that  $\lambda_1\le
\alpha$. As we know, for $\beta\ne1$ it shapes the whole moment
polytope, see Remark \ref{extra_syst} to Theorem \ref{mlt_clmn}.
Here we elucidate the remaining case $\nu=[N-1,1]$, and thus solve
the pure $\nu$-representability problem for all two-row diagrams.
The result can not be extended to three-row diagrams, nor even to
three fermion systems, where  the number of independent
inequalities {\it increases\/} with the rank, see Corollary
\ref{no_fnt} below.
For convenience and a future reference we collect in the next
theorem all known facts.
\begin{theorem}\label{two_row_thm}
For a system $\mathcal{H}_r^\nu$ of rank $r\ge 3$ with two-row
diagram $ \nu=[\alpha,\beta]$, $\alpha+\beta=N$
all constraints on the occupation numbers of a pure state are
given by the following conditions
\begin{enumerate}
\item Basic inequality $\lambda_1\le \alpha$ for $\beta\ne1$.
\item Inequality $\lambda_1-\lambda_2\le N-2$ for $\nu=[N-1,1]$,
$N>3$.\label{excp2row}
\item Inequalities $\lambda_1-\lambda_2\le 1$, $\lambda_2-\lambda_3\le
1$ for $\nu=[2,1]$.\label{excp21}
\item Even degeneracy $\lambda_{2i-1}=\lambda_{2i}$ for
$\nu=[1,1]$.
\end{enumerate}
\end{theorem}
\proof We have already addressed the cases 1 and 4 in Remark
\ref{extra_syst} and Introduction respectively.
\bigskip

\noindent {\it Case 2: Necessity. \/} To prove the inequality
$\lambda_1-\lambda_2\le N-2$ we have to put it into the form of
Theorem \ref{nu thm}
\begin{equation}\label{nu}\sum_i
a_{i}\lambda_{v(i)}\le \sum_k  a^\nu_{k}\mu_{w(k)}.
\end{equation}
This suggests the test spectrum $a=(1,0,0,\ldots,0,-1)$ and the
shortest permutation $v$ that transforms it into
$(1,-1,0,0,\ldots,0)$, which is the cyclic one $v=(2,3,4,\ldots,r)$.
Thus we get the left hand side of the inequality.
To interpret its right hand side $N-2$, notice that
the spectrum $a^\nu$ starts with the
terms
$$a^\nu=(\underbrace{N-1,N-1,\ldots,N-1}_{r-2},N-2,\ldots),$$
corresponding to semi-standard tableaux $T$ with first row of
ones and 
the indices $2,3,\ldots,r$ filling  the unique place in the second
row. Since for pure state $\mu=(1,0,0,\ldots,0)$, then the
shortest permutation $w$ that produces $N-2$ in the right hand
side of (\ref{nu}) is also cyclic $w=(1,2,3,\ldots,r-1)$. The
corresponding Schubert polynomial is just the monomial
$$S_w(x^\nu)=x^\nu_1x^\nu_2\cdots x_{r-2}^\nu.$$
This is a special case of Grassmann permutations discussed in the
next $n^\circ$~\ref{Grass}. Specialization $x^\nu_k\mapsto x_T$ of
Theorem \ref{nu_coeff} transforms it into the product
\begin{equation*}
P(x)=\prod_{i=2}^{r-1}[(N-1)x_1+x_i].
\end{equation*}  Taking the reduced decomposition $v=s_2s_3\cdots
s_{r-1}$ we infer
$$c_w^v(a)=\partial_v
P(x)=\partial_2\partial_3\cdots\partial_{r-1}P(x).$$ The right
hand side is a constant, and the operators $\partial_i$ do not
touch $x_1$. Hence we can put $x_1=0$, that gives
$$c_w^v(a)=\partial_2\partial_3\cdots\partial_{r-1}(x_2x_3\cdots x_{r-1})=1.$$
Since $c_w^v(a)\ne0$, the inequality follows from Theorem \ref{nu
thm}.

\bigskip
\noindent {\it Case 2: Sufficiency. \/} By the convexity it is enough to
construct  {\it extremal states\/} whose occupation numbers are
vertices of the polytope cut out from the  Weyl chamber by the
inequality $\lambda_1-\lambda_2\le N-2$ and the normalization
$\mathop{\mathrm{Tr}}\lambda=N$. The vertices are given first of
all by the fundamental weights normalized to trace $N$
$$\omega_k=(\underbrace{N/k,N/k,\ldots,N/k}_k,0,0,\ldots,0)$$
that generate the edges of the Weyl chamber, except for $\omega_1$
forbidden by the constraint $\lambda_1-\lambda_2\le N-2$. The
latter is replaced by the intersections $\tau_k$ of segments
$[\omega_1,\omega_k]$ with the hyperplane $\lambda_1-\lambda_2=
N-2$
$$\tau_k=(\underbrace{N-2+2/k,2/k,\ldots,2/k}_{k},0,0,\ldots,0).$$
Here we tacitly  assume  that $N>3$, since otherwise $\omega_2$
would be also forbidden. The same condition ensures that the
system $\mathcal{H}_k^\nu$ is non-exceptional for $k\ge 2$, hence
$\omega_k$ are occupation numbers of some pure states by
Proposition~\ref{non_excep}.

To deal with the remaining vertices $\tau_k$ we invoke the
Dadok-Kac construction $n^\circ$~\ref{Dad_Kac} and observe that
the state
$$\psi_k=\text{\scriptsize $\young(1kk\cdot\cdot\cdot k,k)$}+
\frac{1}{\sqrt 2}\sum_{2\le i<
k}\text{\scriptsize$\young(iik\cdot\cdot\cdot k,k)$}$$ has a
disconnected support and the occupation numbers $\tau_k$, $k\ge2$.
Here for clarity we write tableau $T$ instead of the weight vector
$e_T$ and skip an overall normalization factor.

\bigskip
\noindent {\it Case 3. \/}
Here we only briefly sketch the proof that follows a similar
scheme. The second inequality in the form $\lambda_2-\lambda_3\le
N-2$ holds for all $N$, but it becomes redundant for $N>3$. It can
be deduced from Theorem \ref{nu thm}  by calculation of the
coefficient $c_w^v(a)$ for the same $a$ and $w$ as above, but with
another permutation $v=(1,2)(3,4,\ldots,r)$. Then, keeping the
notations of Case 2, we get
\begin{eqnarray*}
c_w^v(a)&=&\partial_3\partial_4\cdots\partial_{r-1}\partial_1P(x_1,x_2,\ldots,x_{r-1})\\
&=&
\partial_3\partial_4\cdots\partial_{r-1}
\frac{P(x_1,x_2,\ldots,x_{r-1})-P(x_2,x_1,\ldots,x_{r-1})}{x_1-x_2}.
\end{eqnarray*}
The operators $\partial_k$, $k\ge3$ do not affect variables
$x_1,x_2$. Therefore we can pass in the fraction to the limit
$x_1,x_2\rightarrow 0$ equal to $(N-2)x_3x_4\cdots x_{r-1}$, that
gives $c_w^v(a)=N-2\ne 0$.

To prove sufficiency of the above inequalities we again have to
look at the vertices of a polytope cut out of the Weyl chamber by
the constraints $\lambda_1-\lambda_2\le 1$,
$\lambda_2-\lambda_3\le 1$, $\mathop{\mathrm{Tr}}\lambda=3$. This
time, along with $\omega_k,k\ge 3$ and  $\tau_k,k\ge 2$, there are
vertices of another type
$$\eta_k=(\underbrace{1+1/k,1+1/k,1/k,1/k,\ldots,1/k}_k,0,0,\ldots,0)$$
for $k\ge3$. They represent occupation numbers of the following
states with disconnected support
$$ \psi_k=\sqrt{k+1}\;\text{\scriptsize $\young(11,2)$}+
\sqrt{2}\;\text{\scriptsize $\young(22,3)$}+ \sum_{3<i\le
k}\;\text{\scriptsize $\young(2i,i)$}\;.\qquad\qed$$

\begin{remark} Two-row diagrams naturally appear in description of
bosonic systems, like photons where polarization plays r\^{o}le of
spin. Representation with diagram {\tiny $\yng(2,1)$} can be
applied both for
bosons and fermions. In this case we calculated all constraints on
the spin and orbital occupation numbers for  small ranks, see
$n^\circ$~\ref{spin_orb_ex}. It appears that the constraints are
 stable and independent of the rank.
\end{remark}

\subsection{Grassmann inequalities}\label{Grass}
Let's return back to the initial pure $N$-represent\-ability problem
for system $\wedge^N\mathcal{H}_r$ and consider a
constraint on its occupation numbers with 0/1 coefficients
\begin{equation}\label{Gr_inq}
\lambda_{i_1}+\lambda_{i_2}+\cdots+\lambda_{i_p}\le b,
\end{equation}
called {\it Grassmann inequality\/}.
For example, all constraints (\ref{rank7}) for system
$\wedge^3\mathcal{H}_7$ are Grassmannian. We assume  that the
Grassmann inequality is {\it essential\/}, meaning that it defines
a  facet of the moment polytope.
Then it should fit into the form of Theorem \ref{nu thm} with
$$a=(\underbrace{1,1,\ldots,1}_p,0,0,\ldots,0)$$
and the {\it Grassmann permutation\/} or {\it shuffle\/}
\begin{equation}\label{Gr_prm}
v=[i_1,i_2,\ldots,i_p,j_1,j_2,\ldots,j_q]:=[I,J],
\end{equation} where $I$ and $J$ are
increasing sequences of lengths $p$ and $q$, $p+q=r$. This is the
shortest permutation
that produces the left hand side of inequality (\ref{Gr_inq}).
Our terminology stems from the observation that for the test
spectrum $a$ the flag variety $\mathcal{F}_a(\mathcal{H})$ reduces
to the {\it Grassmannian\/} $\mathrm{Gr}_p^q(\mathcal{H})$
consisting of all subspaces in $\mathcal{H}$ of dimension $p$ and
codimension $q$.

It is instructive to think about Grassmann permutation $v=[I,J]$
 geometrically as a path $\Gamma$
connecting $SW$ and $NE$ corners of $p\times q$ rectangle, with
$k$-th unit step running to the North for $k\in I$ and to the East
for $k\in J$. The path cuts out of the rectangle a Young diagram
$\gamma$ at its $NW$ corner. We'll refer to $I$ and $J$ as the
{\it vertical\/} and {\it horizontal\/} sequences of the diagram
$\gamma\subset p\times q$ and denote the corresponding shuffle by
$v_\gamma=[I,J]$.
The length of the shuffle $v_\gamma$ is equal to the size $|\gamma|$
of the diagram $\gamma$ and its Schubert polynomial reduces to the
much better understood Schur function
$$S_{v_\gamma}(x)=S_\gamma(x_1,x_2,\ldots,x_p).$$
Observe that $\gamma_{p-k+1}=i_k-k$, and the size of the Young
diagram $\gamma$ related to its vertical sequence by the equation
\begin{equation}\label{vert_size}
|\gamma|=\sum_{1\le k\le p} (i_k-k).
\end{equation}
To get the strongest inequality (\ref{Gr_inq}) we chose $w$ to be
cyclic\footnote{ Actually $w$ is always cyclic for an essential pure
$\nu$-representability inequality. We'll address this issue
elsewhere.} permutation
$$w=(1,2,\ldots, \ell+1)=[2,3,\ldots,\ell+1,1,\ell+2,\ell+3\ldots,r]$$
of length $\ell=\ell(v)=|\gamma|$ for which the right hand side
$b=(\wedge^Na)_{\ell+1}$ of (\ref{nu})  is minimal and equal to
$\ell+1$-th term of the non-increasing sequence
$$\wedge^Na=
\{a_K:=a_{k_1}+a_{k_2}+\cdots+a_{k_N}\mid 1\le k_1<k_2<\cdots<k_N\le
r\}^\downarrow.$$ The sequence consists of nonnegative numbers $m$
each taken with multiplicity
\begin{equation*}
\binom{p}{m}\binom{q}{N-m}.
\end{equation*} Recall that $w$ also
should be the minimal representative in its left coset modulo
stabilizer of $\wedge^Na$.
For the cyclic permutation this amounts to the inequality
$(\wedge^N a)_{\ell}>(\wedge^N a)_{\ell+1}=b$, which tells that
the first $\ell$ terms of $\wedge^Na$ contain all the components
bigger than $b$. The number of such terms is bounded by the
inequality
\begin{equation}\label{mult}
\sum_{m>b}\binom{p}{m}\binom{q}{N-m}=\ell=|\gamma|\le pq.
\end{equation}
To avoid sporadic constraints, assume that the inequality we are
looking for is {\it stable\/}, i.e. remains valid for arbitrary big
rank $r$.
Then the left hand side should be linear in $q=r-p$ and the sum
contains at most two terms: $m=N$ and $m=N-1$.
Thus we end up with two possibilities
\begin{enumerate}
\item $b=N-2$, $p=N-1$, $\ell=r-p$, that gives the inequality
\begin{equation}
\lambda_{i_1}+\lambda_{i_2}+\cdots+\lambda_{i_{N-1}}\le N-2,
\end{equation}
with $\sum_k (i_k-k)=r-p$. \smallskip
\item $b=N-1$, $p\ge N$, $\ell=\binom{p}{N}$, that gives the inequality
\begin{equation}
\lambda_{i_1}+\lambda_{i_2}+\cdots+\lambda_{i_p}\le N-1,
\end{equation}
with $\sum_k (i_k-k)=\binom{p}{N}$.
\end{enumerate}
We will refer  to them as  the Grassmann inequalities of the first
and second kind respectively.
For the inequalities of the first kind the sum $\sum_k
(i_k-k)=r-p$ increases with the rank, and  therefore some of the
involved occupation numbers should move away from the head of the
spectrum.
In contrast, the constraints of the second kind deal only with a
few leading occupation numbers that are independent of the rank.
We analyze them below for $p=N+1$ and postpone a more peculiar
first kind to the next section. The final result is that these
inequalities actually hold true with very few exceptions.

The cyclic permutation $w$ is a special type of shuffle with
column Young diagram of height $\ell$. The corresponding Schur
function is just the monomial
$$S_w(y)=y_1y_2\ldots y_\ell.$$
Applying to $S_w$ the specialization of Theorem \ref{nu_coeff} we
arrive at the product
\begin{equation}\label{Chern}
P(x)=\prod_{1\le k_1<k_2<\cdots<k_N\le
p}(x_{k_1}+x_{k_2}+\cdots+x_{k_N})=\sum_{\gamma}c_\gamma
S_\gamma(x_1,x_2,\ldots,x_p).
\end{equation}
Being symmetric, it can be expressed via
Schur functions
and, by Theorem~\ref{nu thm}, each time $S_\gamma(x)$ enters into
the decomposition with nonzero coefficient
$c_\gamma\ne 0$ we get inequality
\begin{equation}\label{kind2}
\lambda_{i_1}+\lambda_{i_2}+\cdots+\lambda_{i_p}\le N-1,
\end{equation}
where $ i_1<i_2<\cdots<i_p$ is the vertical sequence of Young
diagram $\gamma\subset p\times q$, $|\gamma|=\binom{p}{N}$.

The product $P(x)$
represents the top Chern class of the exterior power
$\wedge^N\mathcal{E}_p$ of the tautological bundle $\mathcal{E}_p$
on Grassmannian $\mathrm{Gr}_p^q$ and the decomposition
(\ref{Chern}) has been discussed in this context \cite{Lascoux}.
However, known results are very limited.

\begin{example}\label{N2} For $N=2$ and any $p\ge N$ the product
$$P(x)=\prod_{1\le i<j\le
p}(x_i+x_j)=S_\delta(x_1,x_2,\ldots,x_p)$$ is just Schur function
with triangular Young diagram $\delta=[p-1,p-2,\ldots,0]$, see
\cite{Macdonald95}. This gives for two fermion system
$\wedge^2\mathcal{H}$ the inequality
\begin{equation}\label{odd}
\lambda_1+\lambda_3+\lambda_5+\lambda_7\cdots\le 1,
\end{equation} that, due
to the normalization $\sum_i\lambda_i=2$, degenerates into
equality and implies even degeneracy $\lambda_{2i-1}=\lambda_{2i}$
of the occupation numbers.

On the other hand, for arbitrary $N$ and minimal value $p=N$ we get
$$P(x)=x_1+x_2+\cdots+x_N=S_\Box(x).$$
The vertical sequence of the one-box diagram {\tiny \yng(1)}\,
gives a nontrivial inequality
\begin{equation}\label{inc}
\lambda_1+\lambda_2+\cdots+\lambda_{N-1}+\lambda_{N+1}\le N-1
\end{equation}
that forces  $N$-th electron into $N$-th orbital, when  the
preceding orbitals are fully occupied. We improve it below.
\end{example}

To the rest of this section we focus upon the next case $p=N+1$
that provides an infinite series of inequalities.
Observe that in this setting  a row diagram $\gamma$ of length
$N+1=\binom{p}{N}$ produces a {\it false\/} inequality
\begin{equation}\label{false1}
\lambda_1+\lambda_2+\cdots+\lambda_N+\lambda_{2N+2}\le N-1,\quad(?)
\end{equation}
that fails for a coherent state given by one Slater determinant
$e_1\wedge e_2\wedge\ldots\wedge e_N$. Similarly, the column
inequality
\begin{equation}\label{false2}
\lambda_2+\lambda_3+\ldots+\lambda_{N+2}\le
N-1\qquad\qquad(?)
\end{equation} {\it fails for even\/} $N$.
Indeed,
in this case the system
$\wedge^N\mathcal{H}_{N+2}\subset\wedge^N\mathcal{H}_r$ is
non-exceptional and hence, by Proposition \ref{non_excep}, the
spectrum
$$\lambda=\frac{1}{N+2}(\underbrace{N,N,\ldots,N}_{N+2},0,0\ldots,0)$$
represents legitimate occupation numbers violating the inequality.

Quite unexpectedly,  all the other diagrams produce a valid
constraint. In plain  language the result can be stated as follows.

\begin{theorem}\label{Gr_kind2}
The occupation numbers of $N$-fermion system $\wedge^N\mathcal{H}$
in a pure state satisfy the following constraint 
$$\lambda_{i_1}+\lambda_{i_2}+\cdots+\lambda_{i_{N+1}}\le N-1
$$
each time $\sum_k(i_k-k)=N+1$, except for inequality
{\rm(\ref{false1})} and, for even $N$, inequality {\rm
(\ref{false2})}.
\end{theorem}
\proof For $p=N+1$ the decomposition (\ref{Chern}) takes the form
\begin{eqnarray*}
P(x)&=&\prod_{1\le i\le
N+1}(x_1+x_2+\cdots+\widehat{x_i}+\cdots+x_{N+1})=\prod_{1\le i\le
N+1}(\sigma_1-x_i)\\
&=&\sum_{0\le k\le
N+1}(-1)^k\sigma_1^{N+1-k}\sigma_k=\sum_{\gamma}c_\gamma
S_\gamma(x_1,x_2,\ldots,x_{N+1}),
\end{eqnarray*}
where $\sigma_k(x)=S_{[1^k]}(x)$ are elementary symmetric
functions, or what is the same Schur functions for the column
diagram $[1^k]$.

For Young diagrams $\tau\subset\gamma$ denote by $t(\gamma/\tau)$
the number of standard tableaux of skew shape $\gamma/\tau$. Then
\begin{equation}\label{col}
c_\gamma=\sum_{k\ge0}(-1)^kt(\gamma/[1^k]).
\end{equation}
Indeed, the coefficient at $S_\gamma$ in
$\sigma_1^{N+1-k}\sigma_k=S_{[1]}^{N+1-k}S_{[1^k]}$ is equal to
the number of ways to build  $\gamma$ from the column diagram
$[1^k]$ by adding cells one at a time. Numbering the cells in the
order of their appearance gives a standard tableaux of shape
$\gamma/[1^k]$ that encodes  the whole building process. Thus  the
coefficient is $t(\gamma/[1^k])$ and the equation (\ref{col})
follows.

For a column diagram $\gamma$ we infer from the last equation
$$c_\gamma=\sum_{k=0}^{N+1}(-1)^k=\begin{cases}0,\quad N\equiv 0 \mod 2,\\
1,\quad N\equiv 1 \mod 2.\end{cases}$$ Henceforth we assume that
$\gamma$ is not a column. Let's combine successive even and odd
terms of the sum (\ref{col})
\begin{equation}\label{ev_odd}
c_\gamma=\sum_{i\ge0}[t(\gamma/[1^{2i}])-t(\gamma/[1^{2i+1}])].
\end{equation}
We claim that
\begin{equation}\label{dif}
t(\gamma/[1^{k}])-t(\gamma/[1^{k+1}])=t(\gamma/[2,1^{k-1}]),
\end{equation} where meaningless terms understood as zeros, e.g. the
right hand side for $k=0$.

Indeed, the building process can be described as an extension of the
partially filled tableau

{\scriptsize
$$\young(1\hfil\hfil\hfil\hfil\hfil\hfil,2\hfil\hfil\hfil\hfil,\cdot\hfil\hfil\hfil,\cdot\hfil\hfil\hfil,\cdot\hfil\hfil\hfil,k\hfil\hfil,\hfil\hfil,\hfil\hfil)$$}\\
to a full standard tableau of shape $\gamma$. One can put the number
$k+1$ either just below $k$ or next to $1$. For the first choice the
number of ways to complete the tableau is $t(\gamma/[1^{k+1}])$,
while for another one the number is $t(\gamma/[2,1^{k-1}])$. Hence
$t(\gamma/[1^k])=t(\gamma/[1^{k+1}])+t(\gamma/[2,1^{k-1}])$.

Combining the last two equations we arrive at the following
representation of the coefficient $c_\gamma$ as a sum of nonnegative
terms
\begin{equation}
c_\gamma=\sum_{i>0}t(\gamma/[2,1^{2i-1}]).
\end{equation}
For a row diagram all terms vanish, while otherwise
$t(\gamma/[2,1])\ne 0$.
Hence $c_\gamma> 0$ if the diagram is neither a row nor a column.
The result now follows from Theorem~\ref{nu thm}. \qed

\begin{example}
For $N=3$ the theorem gives four inequalities listed below together
with the corresponding diagrams
\begin{equation}\label{B&D inq}
\begin{array}{rr}
\text{\scriptsize$\yng(1,1,1,1)$}:\quad
\lambda_2+\lambda_3+\lambda_4+\lambda_5\le
2,&\qquad\text{\scriptsize$\yng(2,1,1)$}:\quad
\lambda_1+\lambda_3+\lambda_4+\lambda_6\le2,\\&\\
\vspace{1mm}\text{\scriptsize$\yng(2,2)$}:\quad\lambda_1+\lambda_2+\lambda_5+\lambda_6\le
2,&
\qquad\text{\scriptsize$\yng(3,1)$}:\quad\lambda_1+\lambda_2+\lambda_4+\lambda_7\le2.
\end{array}
\end{equation}
They are valid for arbitrary rank $r$ and give all
constraints on the occupation numbers for $r\le 7$.

Observe  also an improved version of the inequality (\ref{inc})
\begin{equation}\label{impr}
\lambda_1+\lambda_2+\cdots+\lambda_{N-1}+\lambda_{N+1}+\lambda_{2N+1}\le
N-1,
\end{equation}
coming from the diagram $[N,1]$, and another inequality
$$\lambda_2+\lambda_3+\cdots+\lambda_{N+2}\le N-1,$$
originated from a column diagram and valid only for {\it odd\/} $N$.
\end{example}
\begin{remark} We have considered above only Grassmann inequalities of
the lowest  levels $p=N,N+1$. The higher levels provide further
improvements. For example, the inequalities (\ref{inc}) and
(\ref{impr}) are just the first terms of an infinite series
corresponding to increasing values of $p$
\begin{equation}
\lambda_{i_1}+ \lambda_{i_2}+\lambda_{i_3}+\cdots+\lambda_{i_p}\le
N-1,
\end{equation}
where $i_k=k+\binom{k-1}{N-1}$. For $N=2$ this gives the
inequality (\ref{odd}) and the double degeneracy of the occupation
numbers,
while for $N=3$ we get the inequality
$$\lambda_1+\lambda_2+\lambda_4+\lambda_7+\lambda_{11}+\lambda_{16}+\cdots\le 2,$$
where the differences between the successive indices are natural
numbers $1,2,3,4,\ldots$. The details will be given elsewhere.
\end{remark}

\subsection{Grassmann inequalities of the first kind}
Formally we have such an inequality
\begin{equation}\label{kind1}
\lambda_{i_1}+\lambda_{i_2}+\cdots+\lambda_{i_{N-1}}\le N-2
\end{equation}
each time the Schur function $S_\gamma=S_{v_\gamma}$  enters into
the decomposition
\begin{equation}\label{dcmp}
P(x)=\prod_{N\le j\le
r}(x_1+x_2+\cdots+x_{N-1}+x_j)=\sum_{\ell(v)=\ell}c_vS_v(x).
\end{equation}
Here $\gamma$ is a Young diagram of size $\ell=r-N+1$ with the
vertical sequence formed by the
indices in the above inequality, and $v_\gamma$ is the
corresponding shuffle. In contrast to the previous case, the
product is {\it not\/} a symmetric function and its decomposition
into Schubert polynomials is a challenge.

Let's try a simple case of a row diagram that produces the
inequality
\begin{equation}\label{row_inq}
\lambda_1+\lambda_2+\cdots+\lambda_{N-2}+\lambda_r\le N-2.
\end{equation}
A close look shows that it {\it fails\/}  for odd $\ell=r-N+1=2m-1$
for the  spectrum
$$\lambda=(\underbrace{1,1,\ldots,1}_{N-2},\underbrace{1/m,1/m,\ldots,1/m}_{2m})$$
obtained by merging of the occupation numbers of the systems
$\wedge^{N-2}\mathcal{H}_{N-2}$ and $\wedge^2\mathcal{H}_{2m}$,
see Corollary \ref{cor_mlt_clmn} of Theorem \ref{mlt_clmn}.
Neveretheless
\begin{proposition}\label{row_diag} The inequality
{\rm(\ref{row_inq})} holds for even $\ell=r-N+1$. In this case the
Schur function with  a row diagram enters into the decomposition
{\rm(\ref{dcmp})} with unit coefficient.
\end{proposition}
\proof The row diagram $\gamma$ corresponds to the cyclic
permutation
$$v=v_\gamma=(r,r-1,\ldots,N,N-1)=s_{r-1}s_{r-2}\cdots s_{N-1},$$
where $s_i=(i,i+1)$ are transpositions. We have to calculate the
coefficient $c_v$ of the decomposition (\ref{dcmp}) given by the
equation
$$c_v=\partial_vP(x)=\partial_{r-1}\partial_{r-2}\cdots \partial_{N-1}P(x).$$
The operator $\partial_v$ does not affect the variables
$x_i,i<N-1$, so we can set them to zero and deal with the
polynomial
$$P_0(x)=\prod_{N\le i\le r}(x_{N-1}+x_i)=
\sum_{N\le i_1<i_2<\cdots<i_k\le r}
x_{N-1}^{\ell-k}x_{i_1}x_{i_2}\cdots x_{i_k}.$$ We claim that
\begin{equation}\label{mon}
\partial_vx_{N-1}^{\ell-k}x_{i_1}x_{i_2}\cdots x_{i_k}=
\begin{cases}
(-1)^k& \text{ for } i_s=r-k+s,\\
\quad0&\text{ otherwise}.
\end{cases}
\end{equation}
Let start with the second case $i_1\le r-k=\ell+N-k-1$. In the
following calculation we set to zero all variables that are not
affected by the subsequent operators $\partial_j$. With this
convention we get
\begin{equation}\label{conv}
\partial_{i_1-2}\partial_{i_1-3}\cdots\partial_{N-1}x_{N-1}^{\ell-k}x_{i_1}x_{i_2}\cdots
x_{i_k}=x_{i_1-1}^{\ell+N-k-i_1}x_{i_1}x_{i_2}\cdots x_{i_k}.
\end{equation}
 The resulting monomial is divisible by  $s_{i_1-1}$-invariant
factor $x_{i_1-1}x_{i_1}$ that commutes with  operator
$\partial_{i_1-1}$.  Hence  everything vanishes in the next step
as a result of  the action $\partial_{i_1-1}$ and setting
$x_{i_1-1}=0$.

In the case $i_1=r-k+1=\ell+N-k$ the right hand side of (\ref{conv})
is just the product of the last $k$ variables
$x_{r-k+1}x_{r-k+2}\cdots x_r$ and application of the remaining
operators $\partial_j$, $r-k \le j\le r-1 $ gives $(-1)^k$.

Finally, from the equation (\ref{mon}) we infer
\begin{equation}\label{bndry}
c_v=\sum_{0\le k\le \ell}(-1)^k=\begin{cases}1,\quad \ell \text{ is even},\\
0,\quad \ell \text{ is odd},\end{cases}
\end{equation} and the result follows from Theorem \ref{nu thm}. \qed
\begin{remark}\label{two_row}
The inequality (\ref{row_inq}) is most  appealing for $N=3$
\begin{equation}\label{impr_pauli}
\lambda_1+\lambda_r\le 1,
\end{equation} where it supersedes the Pauli
principle $\lambda_1\le 1$ for even $r$. Note that for three
electron system one- and two-point density matrices are
isospectral and therefore the above inequality holds for both of
them. We first came across this result reading paper \cite{G-H},
where the authors observed that if the 2-point density matrix of a
three fermion system in state $\psi\in
\wedge^3\mathcal{H}_r$  has an eigenvalue equal to one, then the
corresponding eigenform $\omega\in\wedge^2\mathcal{H}_r$ can't
have the full rank $r$. This is trivial for odd $r$, since rank of
$\omega$ is always even.  For even rank this follows from
(\ref{impr_pauli}).
Moreover,  in the latter case the  state
$\psi\in\wedge^3\mathcal{H}_r$ itself  has rank less than $r$.
M.B.~Ruskai also conjectured  inequality (\ref{impr_pauli}) in her
analysis of three fermion and three hole systems \cite{Ruskai07}.
\end{remark}
Observe of the following result, anticipated by many experts.
It may appear not so
trivial if compared with
Theorems \ref{mlt_clmn} and \ref{two_row_thm}.
\begin{corollary}\label{no_fnt}  No finite set of inequalities
gives all constraints on occupation numbers of $N$-fermion system
$\wedge^N\mathcal{H}$, $N>1$ of arbitrary big rank.
\end{corollary}
\proof
Indeed, a finite set $Q$ of linear inequalities
$L_\alpha(\lambda)\le b_\alpha$ includes only finitely many
occupation numbers $\lambda_i$, $i< M$. Every inequality that
follows from $Q$ is a nonnegative  combination
of the inequalities from $Q$, the ordering conditions
$\lambda_{i}-\lambda_{i-1}\le0$, and a multiple of the
normalization equation $\sum_{i=1}^r\lambda_i=N$.

Suppose now that the inequality of
Proposition \ref{row_diag} 
\begin{equation}\label{finite}
\lambda_{1}+\lambda_2+\cdots+\lambda_{N-2}+\lambda_{r}\le N-2
\end{equation}
can be deduced from the system $Q$ for some  $r\gg M$ and even
$\ell=r-N+1$. The coefficients at $\lambda_i$   in the left side
for $i\ge M$ should come from the following linear combination
with non-negative coefficients $a_i$
\begin{eqnarray*}
&&a_1(\lambda_2-\lambda_1)+a_{2}(\lambda_{3}-\lambda_{2})+\cdots+
a_{r-1}(\lambda_{r}-\lambda_{r-1})-a_r\lambda_r=\\
&&-\lambda_1a_1+\lambda_{2}(a_{1}-a_2)+\cdots+\lambda_{r-1}(a_{r-2}-a_{r-1})+
\lambda_{r}(a_{r-1}-a_r)
\end{eqnarray*}
amended 
with a multiple of the normalization equation. The  Abel
transformation shown in the second line implies that the
coefficients $a_i$ should form an arithmetical progression
$a_i=ai+b$ for $M\le i<r$, while $a_r=ar+b-1\ge0$.

Suppose now that $a\ge0$. Then the same combination of
inequalities from $Q$ that produces (\ref{finite}) and the same
coefficients $a_i$ for $i<r$ together with $a_r=ar+b\ge0,\quad
a_{r+1}=a(r+1)+b-1\ge0$ would give a {\it false\/} inequality of
rank $r+1$ obtained from (\ref{finite}) by replacing $r\mapsto
r+1$. Recall that the inequality (\ref{finite}) {\it fails\/} for
odd $\ell=r-N+1$. For $a\le 0$  a similar consideration  gives a
false inequality of rank $r-1$.\qed

Proposition \ref{row_diag} can be extended to two-row diagrams
$\gamma=[\ell-k,k]$. For three fermions this leads to the
constraints
\begin{equation}\label{impr_Pauli2}
\lambda_{k+1}+\lambda_{r-k}\le 1,\quad\text{ for }\quad k+1<r-k,
\end{equation}
that prohibit  more than one electron to occupy {\it two\/}
complementary orbitals. It holds both for even and odd $r$ for
$k>0$. The corresponding coefficients $c_\gamma=c(\ell,k)$ of the
decomposition (\ref{dcmp}) satisfy the recurrence relation
$c(\ell,k)=c(\ell-1,k)+c(\ell-1,k-1)$ and
form the left half of the Pascal triangle
$$\begin{array}{ccccccccccccccccccc}
&&&&&&&&&0&&&&&&&&&\\
&&&&&&&&1&&-1&&&&&&&&\\
&&&&&&&0&&0&&0&&&&&&&\\
&&&&&&1&&0&&0&&-1&&&&&&\\
&&&&&0&&1&&0&&-1&&0&&&&&\\
&&&&1&&1&&1&&-1&&-1&&-1&&&&\\
&&&0&&2&&2&&0&&-2&&-2&&0&&&\\
&&1&&2&&4&&2&&-2&&-4&&-2&&-1&&\\
&0&&3&&6&&6&&0&&-6&&-6&&-3&&0&\\
1&&3&&9&&12&&6&&-6&&-12&&-9&&-3&&-1
\end{array}
$$
with apex at $\ell=-1$,  and 0/1  boundary condition for $k=0$ set
by equation (\ref{bndry}).
We return to the Pascal recurrence relation in a more general
framework below, see equation (\ref{recur}).

Observe a zero in the forth line of the Pascal triangle,
corresponding to diagram {\tiny$\yng(1,1)\,$}. In general, a column
diagram should have zero coefficient, because it produces inequality
\begin{equation}\label{col2}
\lambda_1+\lambda_2+\cdots+\widehat{\lambda_{N-\ell}}+\cdots+\lambda_N\le
N-2 \qquad(?)
\end{equation}
that {\it fails\/} for a coherent state given by one Slater
determinant.

It turns out that the Grassmann inequality of the first kind
(\ref{kind1}) holds for all diagrams, except for a column and an
odd row. To wit
\begin{theorem}\label{kind2_thm}
The occupation numbers of $N$-fermion system $\wedge^N\mathcal{H}_
r$ in a pure state satisfy the following constraint
\begin{equation}\label{type1}
\lambda_{i_1}+\lambda_{i_2}+\cdots+\lambda_{i_{N-1}}\le N-2
\end{equation}
each time $\sum_k(i_k-k)=r-N+1$, {\it except\/} for inequality
{\rm(\ref{col2})} and, for odd $\ell=r-N+1$, inequality
{\rm(\ref{row_inq})}.
\end{theorem}
\proof We've to show that Schur function
$S_\gamma(x)=S_{v_\gamma}(x)$ enters into the decomposition
\begin{equation}\label{dcmp1} P_r(x)=\prod_{N\le j\le
r}(x_1+x_2+\cdots+x_{N-1}+x_j)=\sum_{\ell(v)=\ell}c_vS_v(x),
\end{equation}
provided that $\gamma\subset p\times q$ is neither a column nor an
odd row. Here $p=N-1$, $q=\ell=|\gamma|=r-p$.

Note first of all, that the coefficients of this decomposition are
nonnegative for $v\in S_r$ and can be positive only for shuffles
$v=v_\gamma$. The first claim holds in general for the
coefficients $c_v^w(a)$ of Theorem \ref{nu thm}
$$\varphi^*_a(\sigma_w)=\sum_vc_w^v(a)\sigma_v$$
since the cycle $\varphi^{-1}_a(\sigma_w)\subset
\mathcal{F}_a(\mathcal{H}_r)$ is {\it effective\/}. Here $v$ runs
over representatives of minimal length in left coset modulo
stabilizer of $a$. To include all permutations $v\in S_r$ one has to
deal with a small perturbation $\tilde{a}$ that resolves multiple
entries of $a$. However, since
$\varphi_{\tilde{a}}^{-1}(\sigma_w)\subset
\mathcal{F}_{\tilde{a}}(\mathcal{H}_r)$ is pull back of
$\varphi_a^{-1}(\sigma_w)\subset \mathcal{F}_a(\mathcal{H}_r)$ via
natural projection
$\pi:\mathcal{F}_{\tilde{a}}(\mathcal{H}_r)\rightarrow
\mathcal{F}_a(\mathcal{H}_r)$ defined in Remark \ref{indep},  then
decomposition of $\varphi_{\tilde{a}}^{-1}(\sigma_w)$  and
$\varphi_a^{-1}(\sigma_w)$ involve the same Schubert cycles
$\sigma_v$. This prove the second claim. Let's add as a warning,
that the decomposition (\ref{dcmp1}) actually {\it contains\/}
Schubert polynomials $S_v$ with permutations $v\notin S_r$.

The rest of the proof is purely algebraic. We'll proceed by
induction on $r$ keeping $N$ fixed. For the first meaningful case
$r=N+1$, $\ell=2$, as we know, only row diagram
{\scriptsize$\yng(2)$} appears in the decomposition.

Suppose now the induction hypothesis holds for $P_r(x)$, and
consider the next polynomial
\begin{eqnarray}\label{next1}
P_{r+1}(x)&=&(x_1+x_2+\cdots+x_{N-1}+x_{r+1})P_r(x)\nonumber\\
&=&(x_1+x_2+\cdots+x_{N-1}+x_{r+1})\sum_{\ell(v)=\ell}c_vS_v(x).\label{next}
\end{eqnarray}
We can find its Schubert components using a version of Monk's
formula
$$(\alpha_1x_1+\alpha_2x_2+\alpha_3x_3\cdots)S_v(x)=
\sum_{\ell(vt_{ij})=\ell(v)+1}(\alpha_i-\alpha_j)S_{vt_{ij}},$$
where $t_{ij}=(i,j)$, $i<j<\infty$ is a transposition, see \cite[p.
86]{Macdonald91}.
For a typical term of
(\ref{next1}) this gives
\begin{eqnarray}
(x_1+x_2+&\cdots&+x_{N-1}+x_{r+1})S_v\nonumber\\
&=&\sum_{1\le i<N\le j\ne r+1}S_{vt_{ij}}-\sum_{N\le j\ne
r+1}\mathop{\mathrm{sgn}}(r+1-j)S_{vt_{j,r+1}},\label{Monk}
\end{eqnarray}
where the sums include only those transpositions $t$ for which
$\ell(vt)=\ell(v)+1$. We are interested in the terms
$u_\gamma=vt\in S_{r+1}$ that are shuffles coming from a Young
diagram $\gamma\subset p\times(\ell+1)$ of size $\ell+1$. Let's
single out the row diagram for which Proposition \ref{row_diag}
gives the coefficient $c_\gamma$. The remaining shuffles
$u_\gamma$ do not move the last index $r+1$, and therefore
permutation $v=u_\gamma t_{i,j}$ has a bigger length than
$u_\gamma$ for $j\ge r+1$. Hence a non-row Schur component
$S_\gamma$ in (\ref{Monk}) comes from the sum
$$\sum_{1\le i<N\le j\le r}S_{v t_{ij}}$$
for $v=u_\gamma t_{ij}$, $\ell(v)=\ell(u_\gamma)-1=|\gamma|-1$.
Then $v\in S_r$, and $S_v(x)$ enters into decomposition
(\ref{dcmp1}) only for a shuffle $v=v_\tau$. In this case the
relation $v_\tau=u_\gamma t_{ij}$ just means that $\tau$ is
obtained from $\gamma$ by removing a cell. As a result, we arrive
at the recurrence relation
\begin{equation}\label{recur}
c_\gamma=\sum_{\gamma/\tau=\text{cell}}c_\tau,
\end{equation}
that holds for all {\it non-row} diagrams $\gamma$. This implies
that $c_\gamma>0$ if one can obtain  an even row from $\gamma$ by
removing cells one at a time from a non-row diagram. This can be
done for any diagram different from a column or an odd row. The
inequality (\ref{type1}) now follows from Theorem \ref{nu thm}. \qed
\begin{example} For four fermion system $\wedge^4\mathcal{H}_r$ the
theorem gives inequality
$$\lambda_i+\lambda_j+\lambda_k\le 2,$$
that holds for odd rank $r\ge 7$ and  pairwise distinct indices
satisfying equation $i+j+k=r+3$. For even $r$ one has to exclude
the row inequality $\lambda_1+\lambda_2+\lambda_r\le 2$.
\end{example}

For two-row diagrams equation (\ref{recur}) amounts to the Pascal
recurrence relation discussed in Remark \ref{two_row}. In general,
it allows to get an explicit formula for the coefficient
$c_\gamma$ that is surprisingly similar to the one given in the
proof of Theorem \ref{kind2_thm}, where we borrow the notations.
\begin{corollary}
\begin{equation}
c_\gamma=\sum_{k\ge 0}(-1)^k
t(\gamma/[k])=\sum_{i>0}t(\gamma/[2i,1]),
\end{equation}
where the second equality holds for diagrams $\gamma$ different
from rows and columns.
\end{corollary}
\proof Applying equation the (\ref{recur}) recurrently in conjunction
with Proposition \ref{row_diag}  we find out that $c_\gamma$ is
equal to the number of ways to obtain an even row from $\gamma$ by
removing cells one at a time from a non-row diagram. If $\gamma$
is not a row or a column, then the last step in the process will
be $[2i,1]\mapsto [2i]$. Encoding the process by the standard
tableaux, we arrived at the second formula. The first one follows
from the identity
$t(\gamma/[2i,1])=t(\gamma/[2i])-t(\gamma/[2i+1])$, cf. the proof
of Theorem \ref{kind2_thm}, and holds for all diagrams. \qed

\section{Connection with representation theory}\label{repr_thry}

The solution of $\nu$-representability problem suggested by
Theorem \ref{nu thm} is not feasible, except for very small
systems. For example, for  four fermions  $\wedge^4\mathcal{H}_8$
we confront with an immense symmetric group of degree ${8\choose
4}=70$. Besides, listing of the extremal edges for systems of this
size is all but impossible. A representation theoretical
interpretation of the $\nu$-representability discussed below
often allows to  mollify or circumvent these difficulties.

Let's consider a composition of  the Schur functors
$\mathcal{H}\mapsto
\mathcal{H}^\nu$ called a {\it plethysm \/}
\begin{equation}
[\mathcal{H}^\nu]^\mu=\sum_{|\lambda|=|\nu|\cdot|\mu|}
m^\mu_\lambda\mathcal{H}^\lambda.
\end{equation}
It splits into $\mathrm{U}(\mathcal{H})$ irreducible components
$\mathcal{H}^\lambda$ of multiplicity $m^\mu_\lambda$. It is
instructive to treat the diagrams $\lambda$ and $\mu$  as {\em
spectra}. We are interested in their asymptotic behavior
 for $m^\mu_\lambda\ne 0$ and
$|\mu|\rightarrow\infty$. Therefore we normalize them to a fixed
size $\widetilde{\mu}=\mu/|\mu|$,
$\widetilde{\lambda}=\lambda/|\mu|$, so that
$\mathop{\mathrm{Tr}}\widetilde{\mu}=1$ and
$\mathop{\mathrm{Tr}}\widetilde{\lambda}=N=|\nu|$.

\begin{theorem}\label{Mumford} Every time $m_\lambda^\mu\ne0$ the
couple $(\widetilde{\lambda},\widetilde{\mu})$ belongs to the moment
polytope of the system $\mathcal{H}^\nu$, i.e. there exists its
mixed state $\rho^\nu$ of spectrum $\widetilde{\mu}$, with
occupation numbers $\widetilde{\lambda}$.
Moreover every point of the moment polytope is a convex
combination of such spectra
$(\widetilde{\lambda},\widetilde{\mu})$ of a bounded size
$|\mu|\le M<\infty$.\qed
\end{theorem}
The theorem is a special case of Mumford's description of the moment
polytope, see his appendix in \cite{Ness}. It also holds in more
general Berenstein-Sjamaar settings \cite{Ber-Sja}.

\subsection{Practical algorithm}
\label{alg} For a fixed $M$ the convex hull of the spectra
$(\widetilde{\lambda},\widetilde{\mu})$ from Theorem \ref{Mumford}
gives an inner approximation to the moment polytope, while any set
of inequalities of Theorem \ref{nu thm} amounts to its outer
approximation. This suggests the following approach to the mixed
$\nu$-representability problem, which combines both theorems.
\begin{enumerate}{\sf

\item {Find all irreducible components $\mathcal{H}^\lambda\subset
[\mathcal{H}^\nu]^\mu$ for $|\mu|\le M$.}

\item {Calculate the
convex hull of the corresponding spectra
$(\widetilde{\lambda},\widetilde{\mu})$ that gives an inner
approximation $\mathcal{P}^{\mathrm{in}}_M\subset \mathcal{P}$ for
the moment polytope $\mathcal{P}$.}

\item Identify the facets of $\mathcal{P}^{\mathrm{in}}_M$
that are given by the inequalities of Theorem \ref{nu thm}. They
cut out an outer approximation
$\mathcal{P}^{\mathrm{out}}_M\supset\mathcal{P}$.

\item Increase $M$ and continue until
$\mathcal{P}^{\mathrm{in}}_M=\mathcal{P}^{\mathrm{out}}_M$.}
\end{enumerate}
The algorithm became practical by generosity of the authors of
\texttt{LiE} package \cite{LiE}, who made it publicly available.
It allows to handle plethysms efficiently.   We also benefit from
\texttt{Convex} package by Franz \cite{FranzConv}, who apply a
similar approach to the quantum marginal problem for three qutrits
\cite{Franz,Klyachko2004}.

One can incorporate in the algorithm additional constraints on
spectrum of the mixed state $\rho^\nu$.
In many problems this is just a restriction on the rank
$\mathop{\mathrm{rk}}\rho^\nu\le p$, that bounds the number of
rows of $\mu$. For example, a pure state
$\rho^\nu=|\psi\rangle\langle\psi|$ has rank one, the
corresponding diagram $\mu=[m]$ reduces to a row, and the plethysm
amounts to the symmetric power $S^m(\mathcal{H}^\nu)$. More
generally, for spin-orbital occupation numbers of a system of
electrons of total spin $J$, we have to deal with mixed states of
rank $2J+1$, see Corollary~\ref{spin_orb_ineq} to Theorem~\ref{nu
thm}, and respectively with the diagrams $\mu$ of at most that
height.

\subsection{Particle-hole duality}\label{part_hole}
Here is another application of Theorem \ref{Mumford}. Recall, that
we arrived at the $\nu$-representability problem from the
spin-orbital decompositions (\ref{spin_orb}) of
$n^\circ$~\ref{H_lambda}. In this setting the Young diagram $\nu$
comes together with a rectangular frame $r\times s\supset\nu $,
where $r$ and $s$ are dimensions of the orbital and spin spaces
respectively. Let $\nu^*$ be the {\it complementary diagram\/} to
$\nu$ in the frame $r\times s$, that is $\nu^*_i=s-\nu_{r+1-i}$.
One can think about the representation $\mathcal{H}_r^{\nu^*}$ as
describing the {\it holes\/} of the system $\mathcal{H}_r^{\nu}$.
These are dual systems with a natural pairing
$\mathcal{H}_r^\nu\otimes\mathcal{H}_r^{\nu^*}\rightarrow
\mathcal{H}_r^{r\times s}=\det(\mathcal{H}_r)^{\otimes s}$, that can
be extended to a pairing of the plethysms
$[\mathcal{H}_r^\nu]^\mu\otimes[\mathcal{H}_r^{\nu^*}]^\mu\rightarrow
\det(\mathcal{H}_r)^{\otimes sm}$, where $m=|\mu|$. The latter duality
means that if $\mathcal{H}_r^\lambda$ is a component of
$[\mathcal{H}_r^\nu]^\mu$, then $\mathcal{H}_r^{\lambda^*}$ is a
component of $[\mathcal{H}_r^{\nu^*}]^\mu$ of the same
multiplicity. Here $\lambda^*$ is the complementary diagram to
$\lambda\subset r\times sm$. In view of Theorem \ref{Mumford} this
implies
\begin{corollary} The moment polytope of the hole system
$\mathcal{H}_r^{\nu^*}$ is obtained from the moment polytope of
$\mathcal{H}_r^{\nu}$ by the transformation
$(\lambda,\mu)\mapsto(\lambda^*,\mu)$, where
$\lambda_i^*=s-\lambda_{r+1-i}$. \qed
\end{corollary}

\section{Analysis of some small systems}\label{small}
Here we take the challenge to explore {\it all\/} the constraints
on the occupation numbers. This is clearly a mission impossible.
It moves us from a garden of the carefully selected species we
dealt with in the preceding sections, into the midst of a wild
jungle with no order or end in sight.

To succeed  in this environment we try  the algorithm
$n^\circ$~\ref{alg} first.
However, due to computer limitation, it can be accomplished only
for very small systems. For the pure $N$-representability problem
these are
the systems for which Borland and Dennis
made their prophesy
35 yeas ago \cite{Borland Dennis}.
To move further
we use any tool available, from a
clever guess to a numerical optimization.
The final outcome of this endeavour
are all the constraints for the systems of rank not exceeding 10.
For $r\le 8$ we provide a rigorous proof below. We also have a
proof for system $\wedge^3\mathcal{H}_9$ based on other ideas, not
discussed here. For the remaining cases
the constraints are complete only {\it beyond a reasonable doubt\/}.
To resolve the doubt one has to verify independently that the
vertices of the constructed polytope are legitimate occupation
numbers. We did this using a variety of methods for most of the
vertices, but some still evaded all the efforts. For the latter we
resort to the numerical optimization to check that
they indeed can be approached  very closely within the moment
polytope. The biggest system we treated $\wedge^5\mathcal{H}_{10}$
is bounded by $161$ inequalities.

We are ready to bet a bottle of decent wine for every additional
essential constraint found.

\subsection{Spin and orbital occupation numbers} \label{spin_orb_ex}
Let's start with a simple example of constraints on spin $\mu$ and
orbital $\lambda$ occupation numbers for a system of three
electrons of the total spin $J=1/2$.
By Corollary~\ref{spin_orb_ineq} to Theorem~\ref{nu thm} the problem
is equivalent to mixed $\nu$-representability for $\nu=\text{{\tiny
$\yng(2,1)$}}\,$ and $\mathop{\mathrm{Spec}}\rho^\nu=(\mu_1,\mu_2)$.
A calculation based on the algorithm $n^\circ$~\ref{alg} shows that
the constraints amounts to 5 inequalities
\begin{eqnarray*}
\lambda_1-\lambda_2\le 1+\mu_2,\quad \lambda_2-\lambda_3\le
1+\mu_2,\quad \lambda_1-\lambda_3\le2-\mu_2\\
\lambda_1-\lambda_2-\lambda_3\le 1,\quad
2\lambda_1-\lambda_2+\lambda_4\le 4-\mu_2,\qquad
\end{eqnarray*}
that apparently are independent of the rank. We test them for
$r=4,5$.
Recall that $\lambda$ and $\mu$ are arranged in the non-increasing
order and are normalized to the traces $3$ and $1$ respectively.

\subsection{Pure $N$-representability}\label{Pure N-rep}
The known solution for two fermions, together with the
particle-hole duality $n^\circ$~\ref{part_hole}, bound the pure
$N-$representability problem to the range $3\le N\le r/2$. For
rank $r\le 8$ this leaves us with systems $\wedge^3\mathcal{H}_6$,
$\wedge^3\mathcal{H}_7$, $\wedge^3\mathcal{H}_8$, and
$\wedge^4\mathcal{H}_8$.

For three of them $\wedge^3\mathcal{H}_6$, $\wedge^3\mathcal{H}_7$
and $\wedge^4\mathcal{H}_8$  the algorithm $n^\circ$~\ref{alg} runs
flawlessly and terminates at $M= 4, 8, 10$, respectively. The
independent
constraints grouped by the test spectra $a$,
together with the coefficients $c_w^v(a)$, and cycle decomposition
of the permutations $v,w$ are given in Tables
\ref{Permut_Sys_3x6}--\ref{Permut_Sys_4x8}.

The remaining system $\wedge^3\mathcal{H}_8$ is much harder to
resolve.

\begin{table}
\begin{center}
\item[]\begin{tabular}{llll} \hline\rule{0pt}{4mm}
Inequalities & $v\in S_6$   & $w\in S_{20}$  & $c_w^v(a)$ \\
\hline\rule{0pt}{4mm}
\hspace{-2.5pt}$\lambda_1+\lambda_6\le1$&$(2\;6\;5\;4\;3)$ &  & 1\\
$\lambda_2+\lambda_5\le1$&$(1\;2\;5\;4\;3)$ & $(1\;2\;3\;4\;5)$ & 1\\
$\lambda_3+\lambda_4\le1$&$(1\;3)(2\;4)   $ &   & 1\\
\hline\rule{0pt}{4mm}
\hspace{-11pt}$\quad \lambda_4\le \lambda_5+\lambda_6$ & $(1\;4\;3\;2)$ & $(1\;2\;3\;4)$  & 1\\
\hline
\end{tabular}\end{center}
\caption{\label{Permut_Sys_3x6} 
$N$-representability inequalities for system
$\wedge^3\mathcal{H}_6$.
}
\end{table}

\begin{table}
\begin{center}
\item[]\begin{tabular}{llll}\hline\rule{0pt}{4mm} Inequalities&
$v\in S_7$ & $w\in S_{35}$      & $c_w^v(a)$
\\\hline\rule{0pt}{4mm}
\hspace{-2.5pt}$\lambda_2+\lambda_3+\lambda_4+\lambda_5\le 2$&$(1\;2\;3\;4\;5)$ &  & 1\\
$\lambda_1+\lambda_3+\lambda_4+\lambda_6\le 2$&$(2\;3\;4\;6\;5)$ & $(1\;2\;3\;4\;5)$ & 1\\
$\lambda_1+\lambda_2+\lambda_4+\lambda_7\le 2$&$(3\;4\;7\;6\;5)$ &
& 1\\$\lambda_1+\lambda_2+\lambda_5+\lambda_6\le 2$&$(3\;5)(4\;6)   $ &   & 1\\
\hline
\end{tabular}\end{center}
\caption{\label{Permut_Sys_3x7} 
$N$-representability inequalities for system
$\wedge^3\mathcal{H}_7$.
}
\end{table}

\begin{table}
\begin{center}
\item[]\begin{tabular}{llll}\hline\rule{0pt}{4mm} Inequalities &
$v\in S_8$ & $w\in S_{70}$ & $c_w^v(a)$\\\hline\rule{0pt}{4mm}
\hspace{-2.5pt}$\lambda_1\le1$&$(1)$ &$(1)$& 1\\
\hline\rule{0pt}{4mm}
\hspace{-2.5pt}$\lambda_5-\lambda_6-\lambda_7-\lambda_8\le0$&$(1\;5\;4\;3\;2)$ & & 1\\
$\lambda_1-\lambda_2-\lambda_7-\lambda_8\le0$&$(2\;3\;4\;5\;6)$ & &1\\
$\lambda_1-\lambda_3-\lambda_6-\lambda_8\le0$& $(3\;4\;5\;7\;6)$&  &1 \\
$\lambda_1-\lambda_4-\lambda_6-\lambda_7\le0$&$(4\;5\;8\;7\;6)$ & $(1\;2\;3\;4\;5)$ & 1\\
$\lambda_1-\lambda_4-\lambda_5-\lambda_8\le0$&$(4\;6)(5\;7)$ & &1\\
$\lambda_3-\lambda_4-\lambda_7-\lambda_8\le0$&$(1\;3\;2)(4\;5\;6)$ & &1\\
$\lambda_2-\lambda_4-\lambda_6-\lambda_8\le0$ & $(1\;2)(4\;5\;7\;6)$ & & 1\\
\hline\rule{0pt}{4mm}
\hspace{-2.5pt}$\lambda_2+\lambda_3+\lambda_5-\lambda_8\le2$&$(1\;2\;3\;5\;4)$ & & 1\\
$\lambda_1+\lambda_3+\lambda_6-\lambda_8\le2$&$(2\;3\;6\;5\;4)$& &1 \\
$\lambda_1+\lambda_2+\lambda_7-\lambda_8\le2$&$(3\;7\;6\;5\;4)$ & &1\\
$\lambda_1+\lambda_2+\lambda_3-\lambda_4\le2$&$(4\;5\;6\;7\;8)$& $(1\;2\;3\;4\;5)$ & 1\\
$\lambda_1+\lambda_4+\lambda_5-\lambda_8\le2$&$(2\;4)(3\;5)$ & &1\\
$\lambda_1+\lambda_2+\lambda_5-\lambda_6\le2$&$(3\;5\;4)(6\;7\;8)$ & &1\\
$\lambda_1+\lambda_3+\lambda_5-\lambda_7\le2$ & $(2\;3\;5\;4)(7\;8)$ & & 1\\
\hline
\end{tabular}\end{center}
\caption{\label{Permut_Sys_4x8} 
$N$-representability inequalities for system
$\wedge^4\mathcal{H}_8$.
}
\end{table}


\subsubsection {System $\wedge^3\mathcal{H}_8$}
We managed to decompose plethysm $S^m(\wedge^3\mathcal{H}_8)$ up to
degree $m= 24$, but still have had a discrepancy between the inner
and the outer approximations to the moment polytope.
Actually all facets of $\mathcal{P}_{24}^{\mathrm{in}}$, except for
one, fit
Theorem \ref{nu thm}.
For the remaining facet
\begin{equation*}\label{bad ineq} \lambda_1+\lambda_5+\lambda_6\geq 1\quad(?)\end{equation*}
we use a numerical minimization of the linear form
$L(\lambda)=\lambda_1+\lambda_5+\lambda_6$ over all particle
density matrices. It turns out that the form attains its minimum,
equal to $\tfrac{27}{28}$, at the vertex
\begin{equation}\label{extra_vert}\frac{1}{28}(15,15,15,15,6,6,6,6).
\end{equation}
Adding this vertex
gives a polytope $\mathcal{P}$ whose all facets are covered by
Theorem \ref{nu thm}. Thus $\mathcal{P}$ is the genuine moment
polytope for $\wedge^3\mathcal{H}_8$ given by
31 independent inequalities listed
in Table \ref{Permut_Sys_3x8}.
\begin{table}[h]
\begin{center}
\item[]\small{\begin{tabular}{llll}\hline\rule{0pt}{4mm}
   Inequalities & $v\in S_8$  & $w\in S_{56}$ &$c_w^v(a)$  \\
   \hline\rule{0pt}{4mm}
\hspace{-2.5pt}$\lambda_2+\lambda_3+\lambda_4+\lambda_5\le2$   & $(1\;2\;3\;4\;5)$ &   &1  \\
$\lambda_1+\lambda_2+\lambda_4+\lambda_7\le2$   & $(3\;4\;7\;6\;5)$ & $(1\;2\;3\;4\;5)$  &1  \\
$\lambda_1+\lambda_3+\lambda_4+\lambda_6\le2$   & $(2\;3\;4\;6\;5)$ &   & 1\\
$\lambda_1+\lambda_2+\lambda_5+\lambda_6\le2$   & $(3\;5)(4\;6)$    &  &1  \\
   \hline\rule{0pt}{4mm}
   \hspace{-2.5pt}$\lambda_1+\lambda_2-\lambda_3\le1$   & $( 3\;4\;5\;6\;7\;8)$   &  & 1 \\
$\lambda_2+\lambda_5-\lambda_7\le1$   & $(1\;2\;5\;4\;3)(7\;8)$ &  & 1 \\
$\lambda_1+\lambda_6-\lambda_7\le1$   & $(2\;6\;5\;4\;3)(7\;8)$ & $(1\;2\;3\;4\;5\;6 )$  & 1 \\
$\lambda_2+\lambda_4-\lambda_6\le1$   & $(1\;2\;4\;3)(6\;7\;8)$ &  & 1 \\
$\lambda_1+\lambda_4-\lambda_5\le1$   & $(2\;4\;3)(5\;6\;7\;8)$ &  &1  \\
$\lambda_3+\lambda_4-\lambda_7\le1$   & $(1\;3)(2\;4)(7\;8)$    &  &1  \\
   \hline\rule{0pt}{4mm}
\hspace{-2.5pt}$\lambda_1+\lambda_8\le1$&$(2\;8\;7\;6\;5\;4\;3)$  &$(1\;2\;3\;4\;5\;6\;7)$  & 1 \\
   \hline\rule{0pt}{4mm}
\hspace{-2.5pt}$\lambda_2-\lambda_3-\lambda_6-\lambda_7\le0$   & $(1\;2)(3\;4\;5\;8\;7\;6)$ &   & 1 \\
$\lambda_4-\lambda_5-\lambda_6-\lambda_7\le0$   & $(1\;4\;3\;2)(5\;8\;7\;6)$ &$(1\;2\;3\;4\;5\;6\;7 )$  & 1 \\
$\lambda_1-\lambda_3-\lambda_5-\lambda_7\le0$   &$(3\;4\;6)(5\;8\;7)$  & &1  \\
   \hline\rule{0pt}{4mm}
\hspace{-2.5pt}$\lambda_2+\lambda_3+2\lambda_4-\lambda_5-\lambda_7+\lambda_8\le2$ & $(1\;4\;8\;7\;5)$ &   & 1\\
$\lambda_1+\lambda_3+2\lambda_4-\lambda_5-\lambda_6+\lambda_8\le2$ & $(1\;4\;8\;6\;7\;5\;2)$ & $(1\;2\;3\ldots10\;11 )$ & 1 \\
$\lambda_1+2\lambda_2-\lambda_3+\lambda_4-\lambda_5+\lambda_8\le2$ & $(1\;2)(3\;4\;8\;5\;6\;7)$ & & 1 \\
$\lambda_1+2\lambda_2-\lambda_3+\lambda_5-\lambda_6+\lambda_8\le2$ & $(1\;2)(3\;5\;4\;8\;6\;7)$ & & 1\\
   \hline\rule{0pt}{4mm}
\hspace{-2.5pt}$\lambda_1+\lambda_2-2\lambda_3-\lambda_4-\lambda_5\le0$   & $(3\;6\;4\;7\;5\;8)$ & $(1\;2\;3\ldots11\;12 )$ &1 \\
$\lambda_1-\lambda_2-\lambda_3+\lambda_6-2\lambda_7\le0$   & $(2\;6)(3\;4\;5\;8\;7)$ & & 1 \\
\hline\rule{0pt}{4mm}
\hspace{-2.5pt}$\lambda_1-\lambda_3-\lambda_4-\lambda_5+\lambda_8\le0$   & $(2\;8\;5\;7\;4\;6\;3)$ & $(1\;2\;3\ldots12\;13 )$  & 1 \\
$\lambda_1-\lambda_2-\lambda_3-\lambda_7+\lambda_8\le0$   & $(2\;8\;7\;3\;4\;5\;6)$ &    &1  \\
   \hline\rule{0pt}{4mm}
\hspace{-2.5pt}$2\lambda_1-\lambda_2+\lambda_4-2\lambda_5-\lambda_6+\lambda_8\le1$   & $(2\;4\;3\;8\;5\;7\;6)$ &   &1\\
$\lambda_3+2\lambda_4-2\lambda_5-\lambda_6-\lambda_7+\lambda_8\le1$   & $(1\;4)(2\;3\;8\;5)$ &  & 1 \\
$2\lambda_1-\lambda_2-\lambda_4+\lambda_6-2\lambda_7+\lambda_8\le1$   & $(2\;6)(3\;8\;7\;4)$ & $(1\;2\;3\ldots12\;13 )$ & 1 \\
$2\lambda_1+\lambda_2-2\lambda_3-\lambda_4-\lambda_6+\lambda_8\le1$   &$(3\;8)(4\;5\;7\;6)$  &   & 1 \\
$\lambda_1+2\lambda_2-2\lambda_3-\lambda_5-\lambda_6+\lambda_8\le1$   & $(1\;2)(3\;8)(5\;7\;6)$ &   &1  \\
   \hline\rule{0pt}{4mm}
\hspace{-2.5pt}$2\lambda_1-2\lambda_2-\lambda_3-\lambda_4+\lambda_6-3\lambda_7+\lambda_8\le0$ & $(2\;6\;4\;5\;3\;8\;7)$ & &1  \\
\hspace{-6pt}$-\lambda_1+\lambda_3+2\lambda_4-3\lambda_5-2\lambda_6-\lambda_7+\lambda_8\le0$& $(1\;4\;2\;3\;8\;5)(6\;7)$ & $(1\;2\;3\ldots14\;15 )$  & 1 \\
$2\lambda_1+\lambda_2-3\lambda_3-2\lambda_4-\lambda_5-\lambda_6+\lambda_8\le0$ & $(3\;8)(4\;7)$ &  & 1 \\
$\lambda_1+2\lambda_2-3\lambda_3-\lambda_4-2\lambda_5-\lambda_6+\lambda_8\le0$ & $(1\;2)(3\;8)(4\;7\;5)$ &  &1 \\
\hline
\end{tabular}}\end{center}
\caption{\label{Permut_Sys_3x8} 
$N$-representability inequalities for system
$\wedge^3\mathcal{H}_8$.
}
\end{table}

%


We are actually unhappy with employment  of the numerical
optimization, that can produce no rigorous result. Nevertheless, it
provides a helpful hint about missed vertices.
After some guesses and trials we found the state
\begin{equation*}\label{missed_eqn}
\psi=2[123]+\sqrt{10}[145]+\sqrt{5}[347]+\sqrt{2}\mathbf{[356]}+\sqrt{2}[258]+2[368]+[178],
\end{equation*}
whose occupation numbers give the vertex (\ref{extra_vert}). This
provides a rigorous proof of
the completeness the above constraints. Here $[ijk]=e_i\wedge
e_j\wedge e_k$ is the Slater determinant or, in our general
notations, weight vector $e_T$ corresponding to the semi-standard
tableau $T$ transpose to $[ijk]$. Six triplets $[ijk]$ in the
support of $\psi$, excluding  one $\mathbf{[356]}$ typesetted in
bold face, form a disconnected set. They are remnants of our
failed attempt to produce the missed vertex by the Dadok-Kac
construction $n^\circ$~\ref{Dad_Kac}. Extra tableau
$\mathbf{[356]}$ in the support increases the number of adjustable
parameters, but makes the problem nonlinear. Don't ask how the
coefficients were found.

For those people who don't trust a computer assisted proof we give
an extremal state for every vertex of the moment polytope for the
systems $\wedge^3\mathcal{H}_7$, $\wedge^3\mathcal{H}_8$, and
$\wedge^4\mathcal{H}_8$ listed  in Tables
\ref{Ext_States_Sys_4x8}-\ref{Ext_States_Sys_3x8}. They are
sufficient
for a computer independent proof, provided that one takes for
granted the values of  the coefficients $c_w^v(a)$ in Tables
\ref{Permut_Sys_3x7}--\ref{Permut_Sys_3x8}.
\begin{table}
\begin{center}
\item[]\footnotesize{\begin{tabular}{ll} \hline\rule{0pt}{4mm}
 \qquad\qquad\qquad\qquad\qquad Extremal states  & \qquad\quad{Vertices} \\
 \hline\rule{0pt}{4mm}
  \hspace{-3pt}$ [1234]$ &$(1:1:1:1:0:0:0:0)$ \\
  $ [1234]+[1256]+[3456]$ & $(1:1:1:1:1:1:0:0)$ \\
  $ [1234]+[1256]$ & $(2:2:1:1:1:1:0:0)$ \\
  $ [1234]+[1256]+[1357]+[1467]+[2367]+[2457]+[3456]$ & $(1:1:1:1:1:1:1:0)$ \\
  $ [1234]+[1256]+[1357]+[1467]$ & $(2:1:1:1:1:1:1:0)$ \\
  $ \sqrt{2}[1234]+ [1256]+[1357]+[2367]$ & $(2:2:2:1:1:1:1:0)$ \\
  $ \sqrt{2} [1234]+ [1256]+[1357]+[2457]+[3456]$ & $(2:2:2:2:2:1:1:0)$ \\
  $ \sqrt{3} [1234]+ \sqrt{2} [1256]+ [1357]+[2457]$ & $(3:3:2:2:2:1:1:0)$ \\
  $ \sqrt{2}[1234]+\sqrt{2}[1256]+ [1357]+[1467]+[2367]+[2457]$ & $(3:3:2:2:2:2:2:0)$ \\
  $ \sqrt{2}[1234]+ [1256]+[1357]$ & $(4:3:3:2:2:1:1:0)$ \\
  $ [1234]+[5678]$ & $(1:1:1:1:1:1:1:1)$ \\
  $ \sqrt{2}[1234]+ [1256]+[1278]+[1357]+[1368]$ & $(3:2:2:1:1:1:1:1)$ \\
  $ [1234]+[1256]+[1278]$ & $(3:3:1:1:1:1:1:1)$ \\
  $ \sqrt{3} [1234]+[1256]+[1357]+[1458]+[2358]+[2457]+[3456]$ & $(3:3:3:3:3:1:1:1)$ \\
  $ \sqrt{2}[1234]+\sqrt{2}[1256]+ [1357]+[1368]+[1458]+[1467]$ & $(4:2:2:2:2:2:1:1)$ \\
  $ 2[1234]+\sqrt{2}[1256]+ [1357]+[1458]+[2358]+[2457]$ & $(4:4:3:3:3:1:1:1)$ \\
  $ 2[1234]+\sqrt{2}[1256]+ [1357]+[1368]+[2358]+[2367]$ & $(4:4:4:2:2:2:1:1)$ \\
  $ \sqrt{2} [1234]+[1256]+[1357]+[1458]$ & $(5:3:3:3:3:1:1:1)$ \\
  $ \sqrt{3} [1234]+[1256]+[1357]+[2358]$ & $(5:5:5:3:3:1:1:1)$ \\
  $ [1234]+[1256]+[1278]+[1357]+[1368]+[1458]+[1467]$&$(7:3:3:3:3:3:3:3)$ \\
  $ \sqrt{3} [1234]+ \sqrt{2} [1256]+ [1357]+[1368]$ & $(7:5:5:3:3:3:1:1)$ \\
  $ \sqrt{3}[1234]+ [1256]+[1278]+[1357]+[1368]+[2358]+[2367]$ & $(7:7:7:3:3:3:3:3)$ \\
\hline
\end{tabular}}\end{center}
 \caption{\label{Ext_States_Sys_4x8} Vertices of the moment polytope
of $\wedge^4\mathcal{H}_8$ and the corresponding extremal states.}
\end{table}

\begin{table}
\begin{center}
\item[]\small{\begin{tabular}{ll} \hline\rule{0pt}{4mm}
\qquad\qquad\qquad\qquad Extremal states & \quad\quad Vertices\\
\hline\rule{0pt}{4mm}
 \hspace{-3pt}$[123]$ &$(1\!:\!1\!:\!1\!:\!0\!:\!0\!:\!0\!:\!0\!:\!0) $\\
 $ [123]\!+\![145]$ & $(2\!:\!1\!:\!1\!:\!1\!:\!1\!:\!0\!:\!0\!:\!0)$  \\
 $[123]\!+\![145]\!+\![246]\!+\![356]$ & $(1\!:\!1\!:\!1\!:\!1\!:\!1\!:\!1\!:\!0\!:\!0)$ \\
 $ \sqrt{2}[123]\!+\![145]\!+\![246]$ & $(3\!:\!3\!:\!2\!:\!2\!:\!1\!:\!1\!:\!0\!:\!0) $ \\
 $ [123]\!+\![145]\!+\![167]\!+\![246]\!+\![257]\!+\![347]\!+\![356]$ & $(1\!:\!1\!:\!1\!:\!1\!:\!1\!:\!1\!:\!1\!:\!0) $ \\
 $ \sqrt{2}[123]\!+\![167]\!+\![246]\!+\![257]\!+\![145]$ & $(2\!:\!2\!:\!1\!:\!1\!:\!1\!:\!1\!:\!1\!:\!0) $ \\
 $\sqrt{2}[123]\!+\!\sqrt{2}[145]\!+\![246]\!+\![257]\!+\![347]\!+\![356]$ & $(2\!:\!2\!:\!2\!:\!2\!:\!2\!:\!1\!:\!1\!:\!0) $ \\
 $ [123]\!+\![145]\!+\![167]$ &$(3\!:\!1\!:\!1\!:\!1\!:\!1\!:\!1\!:\!1\!:\!0) $ \\
 $ \sqrt{2} [123]\!+\![145]\!+\![246]\!+\![347]$ & $(3\!:\!3\!:\!3\!:\!3\!:\!1\!:\!1\!:\!1\!:\!0)$ \\
 $ \sqrt{3} [123]\!+\!\sqrt{2} [145]\!+\![246]\!+\![257]$ & $(5\!:\!5\!:\!3\!:\!3\!:\!3\!:\!1\!:\!1\!:\!0) $
 \\\hline\rule{0pt}{4mm}
 \hspace{-3pt}$[178]\!+\![368]\!+\![258]\!+\![567]\!+\![347]\!+\![246]\!+\![145]\!+\![123]$& $ (1\!:\!1\!:\!1\!:\!1\!:\!1\!:\!1\!:\!1\!:\!1)$ \\
$\sqrt{2}[178]\!+\![368]\!+\![567]\!+\![246]\!+\!\sqrt{2}[145]\!+\!\sqrt{2}[123]$& $ (2\!:\!1\!:\!1\!:\!1\!:\!1\!:\!1\!:\!1\!:\!1)$ \\
$\sqrt{2}[178]\!+\![258]\!+\![567]\!+\!\sqrt{2}[246]\!+\![145]\!+\!\sqrt{3}[123]$& $ (2\!:\!2\!:\!1\!:\!1\!:\!1\!:\!1\!:\!1\!:\!1)$ \\
$\sqrt{3}[123]\!+\!\sqrt{3}[145]\!+\![246]\!+\!\sqrt{2}[347]\!+\![356]\!+\!\sqrt{2}[258]$& $ (3\!:\!3\!:\!3\!:\!3\!:\!3\!:\!1\!:\!1\!:\!1)$ \\
$\sqrt{3}[178]\!+\!\sqrt{2}[567]\!+\![347]\!+\![246]\!+\!2[145]\!+\!\sqrt{5}[123]$& $ (4\!:\!2\!:\!2\!:\!2\!:\!2\!:\!2\!:\!1\!:\!1)$ \\
$[178]\!+\![246]\!+\![145]\!+\!\sqrt{2}[123]$& $ (4\!:\!3\!:\!2\!:\!2\!:\!1\!:\!1\!:\!1\!:\!1 )$ \\
$[178]\!+\![258]\!+\![246]\!+\![145]\!+\!\sqrt{2}[123]$& $ (4\!:\!4\!:\!2\!:\!2\!:\!2\!:\!2\!:\!1\!:\!1 )$ \\
$[258]\!+\![567]\!+\![145]\!+\!\sqrt{3}[123]$& $ (4\!:\!4\!:\!3\!:\!3\!:\!1\!:\!1\!:\!1\!:\!1)$ \\
$\sqrt{2}[145]\!+\![246]\!+\![347]\!+\![\mathbf{356}]\!+\!\sqrt{2}[368]$& $ (4\!:\!4\!:\!4\!:\!4\!:\!2\!:\!1\!:\!1\!:\!1]$ \\
$\sqrt{2}[178]\!+\![246]\!+\![145]\!+\!\sqrt{2}[123]$& $ (5\!:\!3\!:\!2\!:\!2\!:\!2\!:\!2\!:\!1\!:\!1 )$ \\
$[368]\!+\![347]\!+\!\sqrt{2}[145]\!+\!\sqrt{3}[123]$& $ (5\!:\!5\!:\!3\!:\!3\!:\!2\!:\!1\!:\!1\!:\!1 )$ \\
$2[123]\!+\!\sqrt{10}[145]\!+\!\sqrt{5}[347]\!+\!\sqrt{2}[\mathbf{356}]\!+\!\sqrt{2}[258]\!+\!2[368]\!+\![178]$& $ (5\!:\!5\!:\!5\!:\!5\!:\!2\!:\!2\!:\!2\!:\!2)$ \\
$[178]\!+\![567]\!+\!\sqrt{2}[145]\!+\!\sqrt{3}[123]$& $ (6\!:\!3\!:\!3\!:\!3\!:\!2\!:\!2\!:\!1\!:\!1 )$ \\
$2[123]\!+\!\sqrt{2}[246]\!+\!\sqrt{3}[\mathbf{356}]\!+\!\sqrt{5}[567]\!+\!2[258]$& $ (6\!:\!5\!:\!5\!:\!5\!:\!2\!:\!2\!:\!1\!:\!1)$ \\
$\sqrt{2}[178]\!+\![258]\!+\!\sqrt{2}[246]\!+\![145]\!+\!\sqrt{3}[123]$& $ (6\!:\!6\!:\!3\!:\!3\!:\!3\!:\!2\!:\!2\!:\!2 )$ \\
$2\sqrt{2}[145]\!+\!\sqrt{2}[246]\!+\!\sqrt{2}[347]\!+\!\sqrt{3}[\mathbf{356}]\!+\!\sqrt{3}[368]$& $ (6\!:\!6\!:\!4\!:\!4\!:\!4\!:\!1\!:\!1\!:\!1)$ \\
$2\sqrt{3}[123]\!+\!\sqrt{6}[145]\!+\!\sqrt{2}[\mathbf{356}]\!+\!2[567]\!+\!\sqrt{3}[258]\!+\!\sqrt{3}[178]$& $ (7\!:\!5\!:\!5\!:\!5\!:\!2\!:\!2\!:\!2\!:\!2)$ \\
$\sqrt{2}[145]\!+\!2[246]\!+\![347]\!+\![\mathbf{356}]\!+\!\sqrt{2}[368]$& $ (7\!:\!7\!:\!4\!:\!4\!:\!4\!:\!2\!:\!1\!:\!1 )$ \\
$\sqrt{3}[246]\!+\!\sqrt{2}[347]\!+\!\sqrt{6}[258]\!+\!2[368]\!+\!2\sqrt{2}[178]\!+\![\mathbf{124}]$&$ (9\!:\!5\!:\!5\!:\!5\!:\!3\!:\!3\!:\!3\!:\!3)$ \\
$\sqrt{3}[258]\!+\![567]\!+\!\sqrt{2}[347]\!+\!\sqrt{2}[246]\!+\!2[123]$& $ (9\!:\!6\!:\!4\!:\!4\!:\!4\!:\!3\!:\!3\!:\!3)$ \\
$3[145]\!+\!\sqrt{6}[246]\!+\!3[347]\!+\!2[\mathbf{356}]\!+\!\sqrt{3}[258]\!+\!\sqrt{14}[368]$& $ (9\!:\!8\!:\!8\!:\!8\!:\!3\!:\!3\!:\!3\!:\!3)$ \\
$\sqrt{2}[178]\!+\![258]\!+\!\sqrt{3}[246]\!+\!\sqrt{2}[145]\!+\!\sqrt{5}[123]$& $ (9\!:\!9\!:\!5\!:\!5\!:\!3\!:\!3\!:\!3\!:\!2)$ \\
$2[123]\!+\!\sqrt{2}[246]\!+\!\sqrt{2}[\mathbf{356}]\!+\!\sqrt{3}[567]\!+\!\sqrt{3}[258]\!+\!\sqrt{2}[368]$& $ (9\!:\!9\!:\!9\!:\!9\!:\!4\!:\!4\!:\!2\!:\!2)$ \\
$2\sqrt{2}[145]\!+\!\sqrt{6}[246]\!+\!\sqrt{6}[347]\!+\!\sqrt{5}[\mathbf{356}]\!+\!\sqrt{2}[258]\!+\!3[368]$& $ (10\!:\!10\!:\!10\!:\!10\!:\!4\!:\!4\!:\!3\!:\!3)$ \\
$\sqrt{5}[178]\!+\![347]\!+\!\sqrt{2}[246]\!+\!\sqrt{2}[145]\!+\!2[123]$& $ (11\!:\!6\!:\!6\!:\!5\!:\!5\!:\!5\!:\!2\!:\!2)$ \\
$\sqrt{3}[178]\!+\![258]\!+\!2[246]\!+\!\sqrt{2}[145]\!+\!\sqrt{6}[123]$& $ (11\!:\!11\!:\!6\!:\!6\!:\!4\!:\!4\!:\!3\!:\!3)$ \\
$\sqrt{3}[178]\!+\!\sqrt{2}[567]\!+\![246]\!+\!2[145]\!+\!\sqrt{5}[123]$& $ (12\!:\!6\!:\!6\!:\!5\!:\!5\!:\!5\!:\!3\!:\!3)$ \\
$[123]\!+\!\sqrt{3}[145]\!+\!2[347]\!+\!2[\mathbf{356}]\!+\!\sqrt{3}[258]\!+\!\sqrt{3}[368]$& $ (12\!:\!12\!:\!7\!:\!7\!:\!4\!:\!4\!:\!4\!:\!4)$ \\
 \hline
\end{tabular}}\end{center}
\caption{\label{Ext_States_Sys_3x8} Vertices of the moment
polytope of $\wedge^3\mathcal{H}_8$ and the corresponding extremal
states. The first ten lines give the same data for
$\wedge^3\mathcal{H}_7$.
}
\end{table}

\subsubsection{Systems of rank 9 and 10}
The results here are less definite. Only for smallest system
$\wedge^3\mathcal{H}_9$ we have a rigorous justification of
completeness for the system of $52$ independent inequalities. For
the next one $\wedge^4\mathcal{H}_9$ we found $60$ constraints,
that give a polytope with 103 vertices. For all of them, except
for two
$$[16,16,16,6,6,6,6,6,6]/21,\qquad [20,14,14,14,14,4,4,4,4]/23,$$
we have proved  rigorously that they belong to the moment
polytope. The remaining two vertices were checked only
numerically. It turns out that the same two vertices would provide
the completeness of $125$ constraints for
$\wedge^4\mathcal{H}_{10}$. The occupation numbers of the
remaining systems $\wedge^3\mathcal{H}_{10}$ and
$\wedge^5\mathcal{H}_{10}$ are bounded by $93$ and $161$
inequalities, but many vertices are still waiting a confirmation
by non-numerical methods.

The facets and vertices of the moment polytopes for all systems of
rank $\le 10$ are available in a computer friendly format at {\sf
http://www.fen.bilkent.edu.tr/ $\sim$
murata/N-Representability.zip}.

\end{document}